%% file: ms.tex
\documentclass{emulateapj}
\usepackage{apjfonts, natbib}

\newcommand{\be}{\begin{equation}}
\newcommand{\ee}{\end{equation}}

\shorttitle{Second results from QUaD}
\shortauthors{QUaD collaboration}

\begin{document}

\slugcomment{Submitted to ApJ}

\title{Second and third season QUaD CMB temperature and polarization power spectra}

\author{
  QUaD collaboration
  --
  C.\,Pryke\altaffilmark{1},
  P.\,Ade\altaffilmark{2},
  J.\,Bock\altaffilmark{3,4},
  M.\,Bowden\altaffilmark{2,5},
  M.\,L.\,Brown\altaffilmark{6,7},
  G.\,Cahill\altaffilmark{8},
  P.\,G.\,Castro\altaffilmark{6,9},
  S.\,Church\altaffilmark{5},
  T.\,Culverhouse\altaffilmark{1},
  R.\,Friedman\altaffilmark{1},
  K.\,Ganga\altaffilmark{10},
  W.\,K.\,Gear\altaffilmark{2},
  S.\,Gupta\altaffilmark{2},
  J.\,Hinderks\altaffilmark{5,11},
  J.\,Kovac\altaffilmark{4},
  A.\,E.\,Lange\altaffilmark{4},
  E.\,Leitch\altaffilmark{3,4},
  S.\,J.\,Melhuish\altaffilmark{2,12},
  Y.\,Memari\altaffilmark{6},
  J.\,A.\,Murphy\altaffilmark{8},
  A.\,Orlando\altaffilmark{2,4}
  R.\,Schwarz\altaffilmark{1},
  C.\,O'\,Sullivan\altaffilmark{8},
  L.\,Piccirillo\altaffilmark{2,12},
  N.\,Rajguru\altaffilmark{2,13},
  B.\,Rusholme\altaffilmark{5,14},
  A.\,N.\,Taylor\altaffilmark{6},
  K.\,L.\,Thompson\altaffilmark{5},
  A.\,H.\,Turner\altaffilmark{2},
  E.\,Y.\,S.\,Wu\altaffilmark{5}
  and
  M.\,Zemcov\altaffilmark{2,3,4}
}

\altaffiltext{1}{Kavli Institute for Cosmological Physics,
  Department of Astronomy \& Astrophysics, Enrico Fermi Institute, University of Chicago,
  5640 South Ellis Avenue, Chicago, IL 60637, USA.}
\altaffiltext{2}{School of Physics and Astronomy, Cardiff University,
  Queen's Buildings, The Parade, Cardiff CF24 3AA, UK.}
\altaffiltext{3}{Jet Propulsion Laboratory, 4800 Oak Grove Dr.,
  Pasadena, CA 91109, USA.}
\altaffiltext{4}{California Institute of Technology, Pasadena, CA
  91125, USA.}
\altaffiltext{5}{Kavli Institute for Particle Astrophysics and
Cosmology and Department of Physics, Stanford University,
382 Via Pueblo Mall, Stanford, CA 94305, USA.}
\altaffiltext{6}{Institute for Astronomy, University of Edinburgh,
  Royal Observatory, Blackford Hill, Edinburgh EH9 3HJ, UK.}
\altaffiltext{7}{{\em Current address}: Cavendish Laboratory,
  University of Cambridge, J.J. Thomson Avenue, Cambridge CB3 OHE, UK.}
\altaffiltext{8}{Department of Experimental Physics,
  National University of Ireland Maynooth, Maynooth, Co. Kildare,
  Ireland.}
\altaffiltext{9}{{\em Current address}: CENTRA, Departamento de F\'{\i}sica,
  Edif\'{\i}cio Ci\^{e}ncia, Piso 4,
  Instituto Superior T\'ecnico - IST, Universidade T\'ecnica de Lisboa,
  Av. Rovisco Pais 1, 1049-001 Lisboa, Portugal.}
\altaffiltext{10}{Laboratoire APC/CNRS, B\^atiment Condorcet,
  10, rue Alice Domon et L\'eonie Duquet, 75205 Paris Cedex 13, France.}
\altaffiltext{11}{{\em Current address}: NASA Goddard Space Flight
  Center, 8800 Greenbelt Road, Greenbelt, Maryland 20771, USA.}
\altaffiltext{12}{{\em Current address}: School of Physics and
  Astronomy, University of
  Manchester, Manchester M13 9PL, UK.}
\altaffiltext{13}{{\em Current address}: Department of Physics and Astronomy, University
  College London, Gower Street, London WC1E 6BT, UK.}
\altaffiltext{14}{{\em Current address}:
  Infrared Processing and Analysis Center,
  California Institute of Technology, Pasadena, CA 91125, USA.}

\begin{abstract}
We report results from the second and third seasons of observation
with the QUaD experiment.
Angular power spectra of the Cosmic Microwave Background are derived
for both temperature and polarization at both 100~GHz and 150~GHz,
and as cross frequency spectra.
All spectra are subjected to an extensive set of jackknife tests to
probe for possible systematic contamination.
For the implemented data cuts and processing technique such contamination
is undetectable.
We analyze the difference map formed between the 100 and 150~GHz
bands and find no evidence of foreground contamination in polarization.
The spectra are then combined to form a single set of results which are
shown to be consistent with the prevailing LCDM model.
The sensitivity of the polarization results is considerably
better than that of any previous experiment ---
for the first time multiple acoustic peaks are detected
in the $E$-mode power spectrum at high significance.
\end{abstract}

\keywords{(cosmology:) cosmic microwave background,
cosmology: observations,
polarization}

\begin{figure*}
\resizebox{\textwidth}{!}{\includegraphics{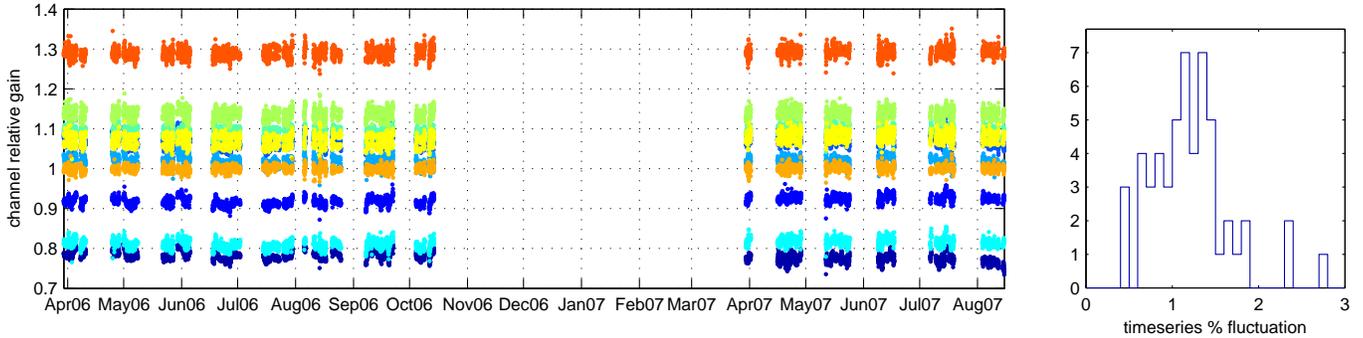}}
\caption{{\it Left:} the relative gains of the first ten 150~GHz detector channels
as measured by elevation nods every half hour over two seasons.
{\it Right:} a histogram of the percentage fluctuations of the timeseries
at left including all channels.}
\label{fig:relgains}
\end{figure*}

\section{Introduction}
\setcounter{footnote}{0}

The anisotropy of the Cosmic Microwave Background (CMB) gives
us direct insight into the structure of the Universe
when it was a tiny fraction of its current age, and is
one of the central pillars of the enormously successful
standard cosmological model.
The temperature anisotropy power spectrum has now been measured
to good precision from the largest angular scales down to
a small fraction of a degree~(e.g.~\cite{reichardt08})
--- the expected series
of acoustic peaks is present and fitting the spectrum
yields tight constraints on the basic parameters of
the cosmological model~(e.g.~\cite{dunkley08}).

The CMB is expected to be polarized at the $\sim10$\% level
principally because of motions in the material at the
time of last scattering.
Since the plasma flows along gradients in the density
field the resulting observable polarization pattern has gradients
($E$-modes), but zero curl ($B$-modes) (e.g.~\cite{hu_white97}).
Given a standard cosmological model fit to the temperature spectrum
($TT$), the $E$-mode spectrum ($EE$), and temperature-$E$-mode cross
spectrum ($TE$), are nearly deterministically predicted ---
only at the largest angular scales is there additional
information.
It is important to remember that, although very successful,
the standard cosmological model (which we will refer to
throughout as LCDM), contains several components which we have
only circumstantial evidence for
(dark matter and dark energy).
Measuring the $EE$ and $TE$ spectra is thus a crucial test of the
overall theoretical paradigm.

As the CMB travels to us through the developing large scale
structure subtle deflections due to gravitational lensing
occur (e.g.~\cite{hu03}).
This converts some fraction of the $E$-mode pattern
into the so-called lensing $B$-modes --- this effect
is most important at smaller angular scales.
In addition, if the cosmogenic theory known as inflation
is correct, there must also be large angular scale $B$-modes
caused by gravity waves propagating through the
primordial plasma (e.g.~\cite{seljakzaldarriaga97}).

The polarization of the CMB was first detected by the
DASI experiment~\citep{kovac02}, and since then
several experiments have reported measurements
of the $EE$ and $TE$ spectra~\citep{barkats05,readhead04,montroy06,
sievers05,page06,bischoff08,nolta08}.
Thus far all reported measurements are consistent with LCDM,
although precision remains limited.
$B$-mode polarization has not yet been detected --- all
results so far are upper limits.
We previously reported preliminary results from QUaD in~\cite{ade07}.
In this paper we report considerably improved results using
data from the second and third seasons of observations.

This paper is structured as follows: in Section~\ref{sec:instobs}
we briefly review the instrument and detail the observations,
Section~\ref{sec:lldataproc}
describes the low level data processing and calibration,
Section~\ref{sec:time2map} outlines the steps used
to make timestream data into maps, Section~\ref{sec:sims}
describes the simulation methodology, Section~\ref{sec:maps2spec}
converts the maps into power spectra, Section~\ref{sec:jackknifes}
gives the results of jackknife tests, Section~\ref{sec:foregrounds}
describes foreground studies, Section~\ref{sec:combspec}
gives the final combined power spectrum results,
Section~\ref{sec:systematics} contains some further investigations
of systematic effects, and
Section~\ref{sec:conclusions} states our conclusions.

\section{Instrument and Observations}
\label{sec:instobs}

The design, implementation, and performance of the QUaD experiment
is described in detail in a companion paper~\citep{hinderks08} hereafter
referred to as the ``Instrument Paper'' --- only a 
very brief summary will be given here.
QUaD was a 2.6~m Cassegrain radio telescope on the mount
originally constructed for the DASI experiment~\citep{leitch02}.
This is an az/el mount with a third axis
allowing the entire optics and receiver to be rotated around
the line of sight (referred to as ``deck'' rotation).
The QUaD receiver consisted of 31~pairs of polarization sensitive
bolometers (PSBs;~\cite{2003SPIE.4855..227J}), 12 at 100~GHz,
and 19 at 150~GHz.
The detector pairs were arranged in two orientation angle groups
separated by 45$^\circ$.
The mount is enclosed in a large bowl shaped reflective ground
shield on top of a tower approximately 1~km from the geographic
South Pole.
The bolometers were read out using AC bias electronics,
and digitized by a 100~Hz, 16~bit ADC.
The raw data were staged on disk at Pole and transferred out daily
via satellite.
QUaD was decommissioned in late 2007.

The observations reported on in this paper were made during the
Austral winter seasons of 2006 and 2007 --- the QUaD telescope was
not able to observe during the summer due to contamination from the Sun.
Complete CMB observation runs occurred on 171 days of 2006 and 118
days of 2007 (defined as an uninterrupted run with the Sun below the horizon).
Of these available 289 days 44 were
rejected after initial low level processing --- mostly due to very bad weather,
with a few due to instrumental problems.
A further 43 days show obvious signs of contamination by the Moon,
and 59 more fail a very conservative Moon proximity cut ---
Moon contamination is discussed in the Instrument Paper and
Section~\ref{sec:mooncontam} below.
This leaves us with a total of 143 days of data which are used in the current analysis.
For simplicity the cut granularity is very coarse ---
we only consider complete days, and if there
is anything wrong with any part of a day we cut the entire day.
It would certainly be possible to include somewhat more data with
additional work.
In Figure~\ref{fig:relgains} we can see the resulting set of days
used --- the monthly gaps are due to the rising and setting
of the Moon.

The QUaD telescope is mounted on a tower at one end of the
MAPO observatory building.
At Pole the celestial sphere rotates about the zenith every 24 hours.
Therefore to minimize the potential for contamination from the building,
or the heat plume from the furnace it contains, each day of
observation starts at a fixed LST such that our chosen
CMB field (centered on RA~5.5h, Dec~-50$^\circ$) has just cleared the 
laboratory building.
The observations are split into two blocks, each of eight hours,
with special calibration observations
before and after each block.
Between the two blocks the entire telescope is rotated by 60$^\circ$ around
the line of sight, and then approximately 30~minutes are allowed for
thermal stabilization.
The total observation schedule takes about 19~hours, with the
remainder of the 24~hour period being taken up by fridge cycling.

Each 8~hour block of CMB observations is divided into
16~half hour periods.
Each starts with an observation of the internal calibration
source followed by an ``elevation nod'' --- the telescope
is moved up and then down again by one degree in elevation
injecting a large signal into the data stream due to the atmospheric
gradient (see the Instrument Paper for details).
The telescope is then scanned back and forth five times over a
7.5$^\circ$ throw in azimuth, with the scan being applied as a
modulation on top of sidereal tracking.
The scan rate is 0.25$^\circ$ per second in azimuth translating
to around 0.16$^\circ$ per second on the sky at our observing elevation.
The pointing position is then stepped by 0.02$^\circ$ in declination
and the process repeated four times.
Including time for moves and settling, these calibration
observations, plus the 40 ``half-scans'' of 30~seconds each,
take half an hour.

During the first half hour of each pair the pointing center
is RA 5.25h.
This is then set to RA 5.75h and the declination offsets
repeated during the second half hour.
The observations are thus taken in a lead-trail manner where
the scanning pattern is identical in ground fixed
azimuth-elevation coordinates between the two half
hours of each pair.
By subtracting the lead and trail data one therefore cancels
any signal coming from the ground which is constant over
the half hour, while producing a difference map of the sky.
We do see significant ground pickup and lead-trail
field differencing is used throughout the analysis presented in
this paper (see Section~\ref{sec:fielddiff}).

\begin{figure}[h]
\resizebox{\columnwidth}{!}{\includegraphics{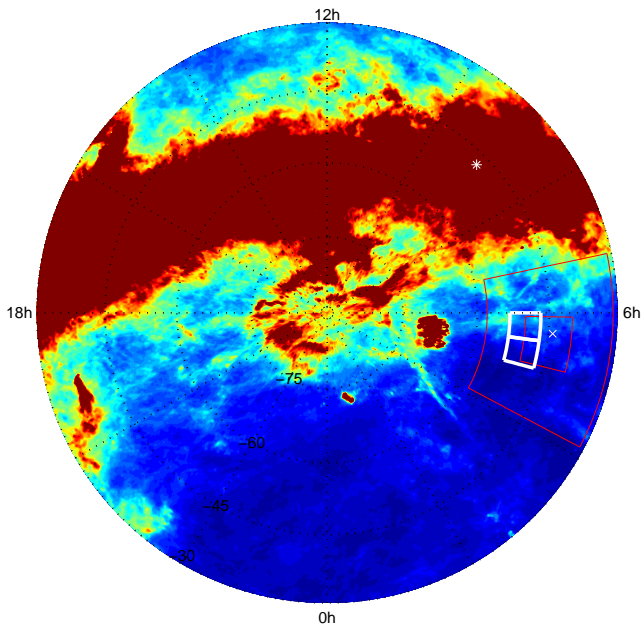}}
\caption{Location of the QUaD lead-trail fields delineated in white
on an equal area azimuthal projection about the SCP.
The color map is the prediction of dust emission intensity
at 150~GHz from FDS model 8~\citep{finkbeiner99}.
The color scale is linear from 0 to 100~$\mu$K,
and is heavily saturated.
(The B03 deep and shallow regions are delineated in red,
while the white asterisk and cross show the locations
of RCW38 and PKS0537-441 respectively.)}
\label{fig:fieldmap}
\end{figure}

From hour to hour the declination offsets
are cumulative, but they are reset before the second
eight hour block.
The telescope therefore scans a ``letter box'' region
0.64$^\circ$ high in Dec twice per day, at two different line of
sight rotation angles.
Each day the observation region steps by 0.64$^\circ$ in Dec to cover
the entire field as rapidly as possibly (in $\approx9$~days), and then cycles around
with a 0.16$^\circ$ offset to generate an eventual four fold
interleaved coverage pattern (after 37~days).
This observation pattern was repeated throughout the 2006 and 2007
seasons a total of $\approx8.5$ times.
Figure~\ref{fig:fieldmap} shows the location of the QUaD
field.

Note that only azimuth scanning is used.
Since the telescope is only $\sim1$~km from the Earth's rotational
axis essentially zero ``cross-linking'' of the map occurs.
For the multipole range presented in this paper this has only
a small negative impact on the final CMB power spectrum
results.

The bright galactic HII region RCW38 was observed on 11 days
distributed through 2006 and 2007, to monitor the beam offset
angles and shapes, and the bright quasar PKS0537-441 was observed
during the 2007 season to further study the beam shapes.
These observations are used in the analysis below
(see Sections~\ref{sec:detoffang} and~\ref{sec:beamwid}).

We also observed several other discrete sources (Cen A, Galactic Center, Moon),
as well as conducted a survey of part of the galactic plane.
These observations will be described in future papers.

\section{Low Level Data Processing}
\label{sec:lldataproc}

This section describes the low level data processing
which occurs before the timestream is binned into
maps (Section~\ref{sec:time2map}) and
analyzed to generate simulations (Section~\ref{sec:sims}).
The data is deconvolved to remove the effects of detector
timeconstants, relative gain calibrated,
and field differenced.

\subsection{Deconvolution of detector temporal response}
\label{sec:deconv}

The initial stages of the data reduction are performed
on each day of data individually.
The first step is deconvolution of the detector time constants ---
the bolometers have non-instantaneous response to changes in
incident optical power, and hence the output data timestream
is a low pass filtered version of the desired input waveform.
With knowledge of the detector time-constants this filtering
can be undone at the price of increased high frequency
noise.

We measured the temporal response of our detectors in situ
using an external Gunn oscillator source as described
in the Instrument Paper.
We find that many of the detectors are well fit by a simple
single time-constant model.
The median primary time constant is 16~ms.
However a substantial fraction ($\sim 50$\%) require a second additive
time-constant to obtain a good fit --- we are hypothesizing
that some fraction of the incident heat goes into a second
reservoir which has a weaker coupling to the thermal bath.
In a couple of cases the second time-constant is several
seconds long but the two time-constant model is still a
good fit --- we retain these detectors.
In two other cases the dual time-constant model is not
a good fit and we reject these detectors leading to the
loss of two channel pairs (both at 100~GHz).

To check that the deconvolution process is working, and
that the timeconstants are stable over time, we examine
forward-backward jackknife maps of the bright
compact source RCW38 taken on 11 days
distributed through 2006 and 2007.
These show no detectable residuals and hence cancellation
to $\ll 1$\%.

As part of the deconvolution process the timestream
is also low pass filtered to $<5$~Hz.
After deconvolution we de-glitch
to remove cosmic ray hits etc.\ (loosing $\ll1$\% of the data).

For this analysis we also exclude two additional channel
pairs; one (at 150~GHz) due to a time evolving scan synchronous
signal, and a second (at 100~GHz) due to strongly
atypical pair differenced noise.
We are thus left with 9 of the possible 12 pairs at 100~GHz,
and 18 of the 19 pairs at 150~GHz.

\subsection{Relative gain calibration via elevation nods}
\label{sec:relgains}

We measure the relative gains of the detector channels
using the elevation nod method mentioned in
Section~\ref{sec:instobs} above.
The airmass through which each channel pair was
looking is calculated from the elevation encoder
reading and then regressed against the observed signal
to yield a calibration factor in volts per airmass.
We then simply normalize each channel's gain to the mean
of all channels:

\begin{equation}
V'_i(t) = V_i(t)  \frac{\overline{g}}{g_i}
\end{equation}

\noindent where subscript $i$ is a loop over channels,
$V(t)$ are the detector timeseries data, and $g$ are
the elevation nod gain factors.
This equalizes the gains both within, and between, channel
pairs.

The nominal accuracy of each elevation nod gain measurement
is $\ll 1$\%.
Weak trends in the apparent relative gains are observed
over short and long timescales, the cause of which is unknown.
However as we see in Figure~\ref{fig:relgains} this leads to
fluctuation over the entire two season time span of only $\approx 1$\% rms.
In this analysis we choose to regard these variations as real
and the relative gains derived from each elevation
nod are applied to the subsequent 40 half-scans.

However even if these apparent variations are false,
random errors of this magnitude will cause negligible 
leakage of total intensity to polarization (hereafter $T$ to pol.\ leakage)
as, for example, $T$ will leak sometimes into $+Q$ and
and sometimes into $-Q$, averaging down in the final maps
over both time and detector pairs.

Possible systematic errors in the relative gains are of much
greater concern and are discussed and simulated further in
Section~\ref{sec:ditherrelgains}.
Imperfect deconvolution would lead to the relative
gain of a detector pair being a function of temporal
frequency.
Since the elevation nods measure the gain at
an effective frequency well below the CMB measurement band
this would result in systematic $T$ to pol.\
leakage.
Using the Gunn oscillator derived time constants
and deconvolution procedure described in Section~\ref{sec:deconv}
the lack of ``monopole'' residuals in pair difference jackknife maps
of the bright source RCW38 indicate that the low
frequency elevation nod gains are accurate to better than 1\%
at high (beam scale) frequencies.

\subsection{Examination of timestream and noise spectra}

\begin{figure}[h]
\resizebox{\columnwidth}{!}{\includegraphics{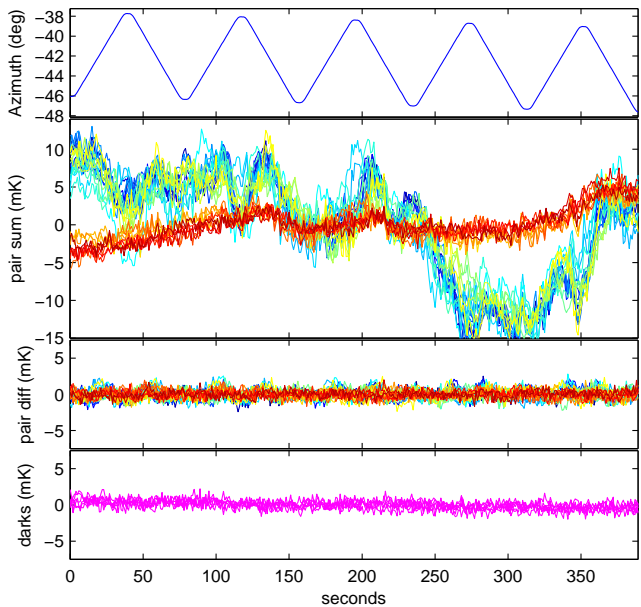}}
\caption{Timestream data for a sample scan set.
The top panel shows the azimuth angle, and the middle panels
the pair sum and difference detector timestreams with relative
gain calibration applied;
red/orange colors are the 100~GHz pairs, while blue/green colors
are the 150~GHz.
The bottom panel shows the dark channels.
In all cases an approximate scaling to temperature units
has been applied.
For the purposes of this illustration the timestreams have been
heavily low pass filtered (to $\leq 1$~Hz).}
\label{fig:timestream}
\end{figure}

The sum of the signals from a PSB pair measures the total
intensity of the incident radiation, while the difference
measures polarization.
Having performed the relative gain calibration and
sum/difference operations Figure~\ref{fig:timestream}
shows a typical scan set at this stage of the processing.
Atmospheric emission at 100 and 150~GHz is dominated by
oxygen and water vapor respectively.
The intensity of the radiation received is an integral over
the temperature and density
along the atmospheric column through which the telescope is looking.
Water vapor is poorly mixed in the atmosphere
leading to much greater fluctuations at 150~GHz.
From the lack of an obvious scan synchronous component
we can infer that the wind blows inhomogeneities through
the beam faster than the telescope scans.
Since the QUaD beams do not diverge until high in the
atmosphere the pair sum timestreams are highly correlated
across the array.
As expected the atmospheric emission is clearly very weakly
polarized leading to strong cancellation in the pair
difference timestreams.

\begin{figure*}[h]
\resizebox{0.5\textwidth}{!}{\includegraphics{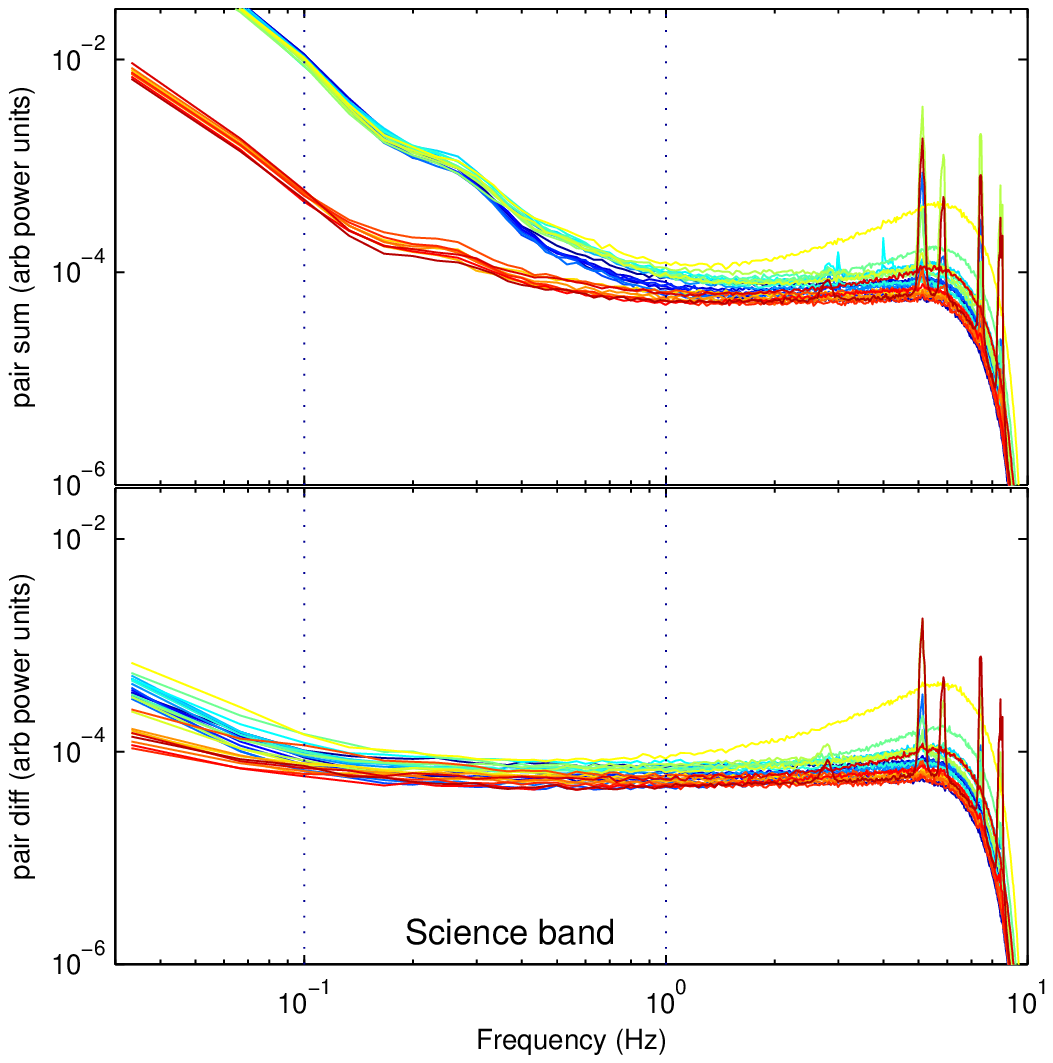}}
\resizebox{0.5\textwidth}{!}{\includegraphics{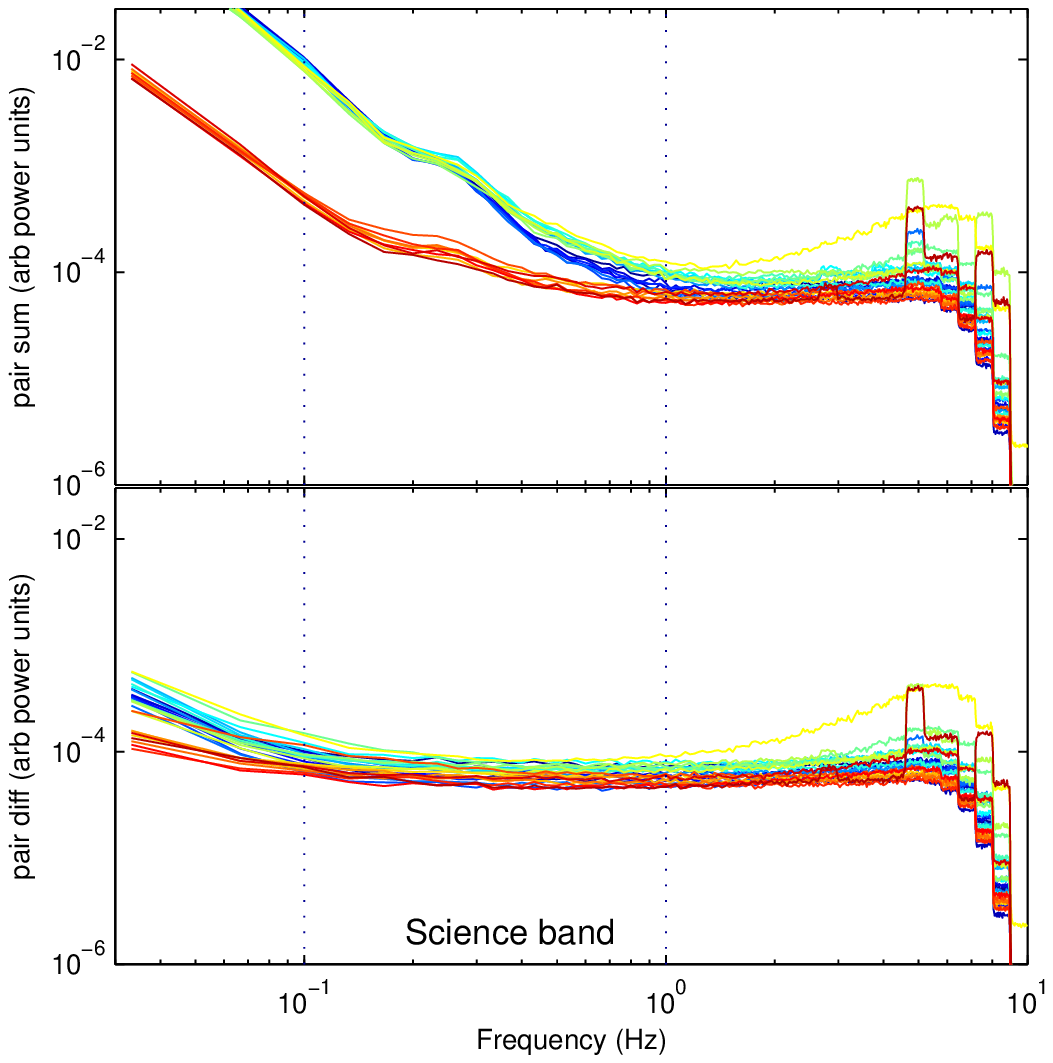}}
\caption{{\it Left:} Timestream PSDs averaged over a single day of data.
Pair sum and pair difference spectra are shown;
red/orange colors are the 100~GHz pairs, while blue/green colors
are the 150~GHz.
The range of frequencies corresponding to $200<\ell<2000$ is enclosed
within the dotted lines.
See the text for further details.
{\it Right:} The same thing for simulated noise timestream
(see Section~\ref{sec:noisims}).}
\label{fig:todspec}
\end{figure*}

Figure~\ref{fig:todspec} shows the mean power spectral
densities (PSDs) of the detector timestreams.
To make this plot spectra were computed for each channel
and half-scan and then averaged over a day of data.
We see substantial $1/f$ noise in the pair sum data.
The $1/f$ knee shifts up in bad weather and
down in good --- the plot is for a day of intermediate quality.
For our scanning speed (0.25$^\circ$/sec in azimuth) and observing
elevation ($\sim50^\circ$) the conversion from timestream
frequency~$f$ to multipole on the sky is $\ell \sim 2000 f$,
giving a ``science band'' of $0.1 < f < 1$~Hz.
Even on the worst days used in this analysis the pair
difference spectra remain close to white within this range.
See the Instrument Paper for further details of the
sensitivity.
Note the lack of narrow line noise within the
science band --- some channels show microphonic lines at much higher
frequencies.
The roll-off above 5~Hz is imposed in the initial processing
as part of the deconvolution procedure.
Also note the uniform noise properties of the
detectors within each frequency group.
(The ``roll up'' towards higher frequencies observed in
some pairs is due to the deconvolution of exceptionally
long time constant detectors.)

\subsection{Long term gain equalization via calibration source}
\label{sec:gainsupcor}

\begin{figure*}[h]
\resizebox{\textwidth}{!}{\includegraphics{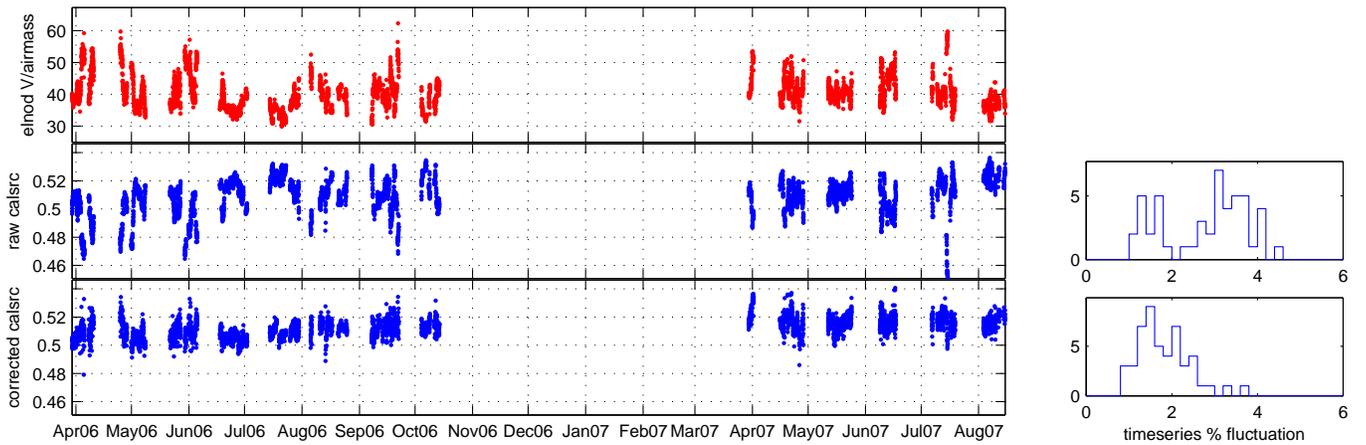}}
\caption{{\it Top:} the atmospheric emission measured by elevation nods in volts
per airmass for a sample detector.
{\it Middle:} the sinusoidal modulation amplitude of the calibration source
as observed by the same channel.
{\it Bottom:} the calibration source amplitude after application
of the loading gain suppression correction (see text).
Histograms over all channels of the percentage fluctuation in the
pre- and post-corrected timeseries are shown to the right.}
\label{fig:gainsupcor}
\end{figure*}

The QUaD telescope was equipped with a battery
powered, remote controlled calibration source mounted
behind the secondary mirror, inside the foam cone.
Before each half hour of observations the 45$^\circ$ flip mirror was commanded
down and the polarizer grid rotated several times
injecting a sinusoidally modulated signal into the detector timestreams ---
see the Instrument Paper for details.
The low level analysis measures the modulation amplitude for each channel.
Figure~\ref{fig:gainsupcor} shows the time series for one channel over the
whole season and also the volts per airmass as measured
by the elevation nods.
There is a clear anti-correlation --- as the atmosphere becomes more
opaque the atmospheric loading goes up, suppressing the detector
gains.
(The source temperature is monitored and shows no correlation with
the external temperature.)
For the good weather data used in this analysis the suppression is 
$\leq10$\%.
We make a linear regression of the mean calibration source timeseries
against the mean elevation nod timeseries within each frequency band and use
this to apply a correction.
After application of this correction we believe the absolute
gain of the QUaD system to be stable at the few percent level
over the entire 2006 and 2007 seasons.

\subsection{Field differencing}
\label{sec:fielddiff}

If we do not field difference ground pickup produces
obvious artifacts in the final co-added maps.
In the timestream this pickup is normally not visible above the detector and
atmospheric noise as we see in Figure~\ref{fig:timestream}.
However on the very worst days, the very worst pairs
can show pickup equivalent to as much as $5$~mK CMB across a half-scan in the
pair differenced data.
We are confident that this is ground signal because
it is not present in the dark channels, is fixed in
azimuth angle, is worst in the bottom row pixels, and correlates
with the amount of snow present on the ground shield ---
see the Instrument Paper for further details.

We note that the ground signal does not show structure
smaller than a few degrees which is expected since it
is near field pickup.
It may be possible to remove the ground signal
sufficiently well using a template based approach,
and we are investigating this option.
However for the analysis presented here we have applied
simple lead-trail differencing throughout.
In doing so we make the assumption that the ground signal,
which clearly does change from day to day, is stable
to the relevant level of accuracy over the half hour timescale.
In the field differenced maps no artifacts are visible.
Ultimately the jackknife tests presented
in Section~\ref{sec:jackknifes} are the most sensitive
test for residual ground contamination.

The start/end points of each half-scan are carefully
tuned to give a best fit to the ideal linear scan motion.
We then apply field differencing point-by-point
in the timestream by subtracting from each half-scan its
partner occurring half an hour later.
The delta RA between the differenced points has a standard
deviation of 0.007$^\circ$. 

Note that for a Gaussian random field, such as the CMB,
the power spectrum of a difference field (in the limit of no correlations) is
twice that of an un-differenced field, with
an associated increase in the sample variance
due to the reduced effective sky area.
The QUAD field differencing is explicitly modeled 
in our analysis pipeline.

\section{From Timestream to Maps}
\label{sec:time2map}

To make polarization maps three more ingredients are required - knowledge
of the overall pointing of the telescope, of the relative pointing
of each of the detectors, and of the polarization angles
and efficiencies of the detectors.

\subsection{Overall telescope pointing}
\label{sec:pointing}

The QUaD mount used a nine parameter online pointing model
derived from optical and radio observations as described in
the Instrument Paper.
During special radio pointing runs this was shown to have
an absolute accuracy over the hemisphere of $\sim0.5'$ rms.

In addition a pointing check was performed on RCW38 before and
after each eight hour block of CMB observations.
These also indicate a $\sim0.5'$ rms wander in the absolute
pointing.
Attempts were made to use the measured offsets to make an offline
pointing correction but it was not possible to demonstrate
any clear improvement, although a few days were rejected due to
abnormally large pointing errors.
Note that the effect of a pointing wander of the observed
magnitude is included in our simulations below.

\subsection{Detector offset angles}
\label{sec:detoffang}

On eleven days throughout the 2006 and 2007 seasons full
day observations were conducted of RCW38
at three line-of-sight orientations (``deck'' angles).
For each day the data from each channel were fit to the six parameter
model of an elliptical Gaussian beam with free centroid positions,
orientation angle and widths.
We did the fits both in the timestream, and in maps, yielding
equivalent results.
The scatter in the centroid positions for a given detector over
the set of days is $\sim0.5'$ consistent with the overall pointing wander
of $\sim0.5'$ rms discussed above.
There is hence no evidence for systematic
changes in the detector offset angles over time.
We therefore take the detector offset angles as the mean over the set
of observed values in the RCW38 runs and estimate their uncertainty
as a negligible $\sim0.15'$.

\subsection{Detector polarization angle and efficiency}
\label{sec:polpar}

Our best measurements of the polarization angles of the
detectors come from in situ measurements of an external
source as described in the Instrument Paper.
A chopped thermal source was placed behind a polarizing grid and
observed with the telescope at many rotation angles,
the signal from each PSB tracing out a sinusoid.
The phase of these sinusoids gives the detector polarization
angle, the fitted values agreeing with the design values 
with a scatter of around one degree rms.
Since this is compatible with the estimated measurement uncertainty
we have used the design values in this analysis when
constructing maps.

The degree to which the sinusoid mentioned above fails
to reach zero represents the response of a measurement channel to
anti-aligned radiation.
This ratio of minimum to maximum response is 
conventionally referred to as the cross polar leakage
$\epsilon$.
Our measured values of $\epsilon$ have a mean of 0.08 
with an rms scatter of 0.015.
In this analysis we have assumed the mean value to apply to all channels
when constructing maps, and included scatter in the
simulations (see Section~\ref{sec:sigsims}).
Note that for an experiment of this type cross polar leakage
does not imply $T$ to pol.\ leakage --- it is simply a small loss
of efficiency, which must be corrected by an additional
calibration factor applied only to the pair difference data.
The effect of systematic error on $\epsilon$ is
discussed in Section~\ref{sec:poleff} below.

\subsection{Map Making}
\label{sec:mapmaking}

To make maps we perform the following operations:
sum and difference the detector timestreams for each pair,
remove a third order polynomial across each 30~second half-scan,
and bin into a grid of pixels weighting by the inverse variance
of each half-scan.
In this analysis the pixelization is in RA and Dec,
and the pixels are 0.02$^\circ$ square.
The polynomial subtraction removes the bulk of the
atmospheric $1/f$ noise allowing the simplicity of
``naive'' map making without incurring a large noise penalty.

For the pair sum data only the signal times the weights, and the
weights themselves, must be accumulated to form the temperature map.
For the pair difference the product of the
data and the sine and cosine of the detector angle as projected
on the sky are accumulated.
Then a 2x2 matrix inversion is performed for each pixel
to produce $Q$ and $U$ maps.
We emphasize that for this matrix to be invertible
any given pixel needs to have been measured at only
two distinct grid angles --- in the presence of noise, angles
separated by 45$^\circ$ are optimal.
The only gain from having a more uniform distribution of angles
is in averaging down a limited number of systematic effects
(such as beam size or pointing mismatch).
In practice since we have detector pairs of two orientation angles, and observe
at two deck angles, pixels in the central region of our
maps have been measured at four angles.

\subsection{Absolute Calibration}
\label{sec:abscal}

At this point we have $T$, $Q$, and $U$ maps at each frequency in raw
detector units (volts).
To scale these into $\mu$K we perform a correlation analysis
versus two noise independent temperature maps from the
2003 flight of the Boomerang experiment (B03) --- which in turn have
been calibrated against WMAP with a 2\% stated uncertainty~\citep{masi06}.
The B03 maps are first passed through the QUaD simulation pipeline
(described in Section~\ref{sec:sigsims} below).
The raw QUaD and ``B03 as seen by QUaD'' maps are then both
apodized by the QUaD sensitivity mask (see Section~\ref{sec:apmask}),
Fourier transformed,
and cross spectra taken between the QUaD map and one B03 map,
and between the two B03 maps.
For each bandpower $b$ we then calculate our absolute calibration
factor as,

\begin{equation}
a_b = \frac{w_{Q,b}}{w_{C,b}} \frac{\left\langle m_R m_C \right\rangle}{\left\langle m_R m_Q \right\rangle}
\label{eqn:abscal}
\end{equation}

\noindent where $w$ is the Fourier transform of the beam, 
$m$ are the modes of the Fourier transform of the apodized map,
$R$, $C$ and $Q$ refer to the ``reference'', ``calibration'' and QUaD
maps respectively, and the mean is taken over the modes
in a given annulus of the Fourier plane. 

This calibration factor should ideally be a constant value for
each bandpower (multipole range).
In practice we find that it is for $200< \ell <800$ where the
B03 beam correction (and therefore uncertainty on that correction)
is modest, and we therefore take the average value across that range.
We have also performed the same operation
using various combinations of the WMAP Q, V and W band maps.
Due to the much larger beams of WMAP, the beam corrections are
very large, and the signal to noise low.
However based on these results, and the point to point scatter
in the B03 analysis, we estimate a 5\% uncertainty
in our primary B03 derived absolute calibration factors.

\subsection{Map Results}

\begin{figure*}[h]
\resizebox{\textwidth}{!}{\includegraphics{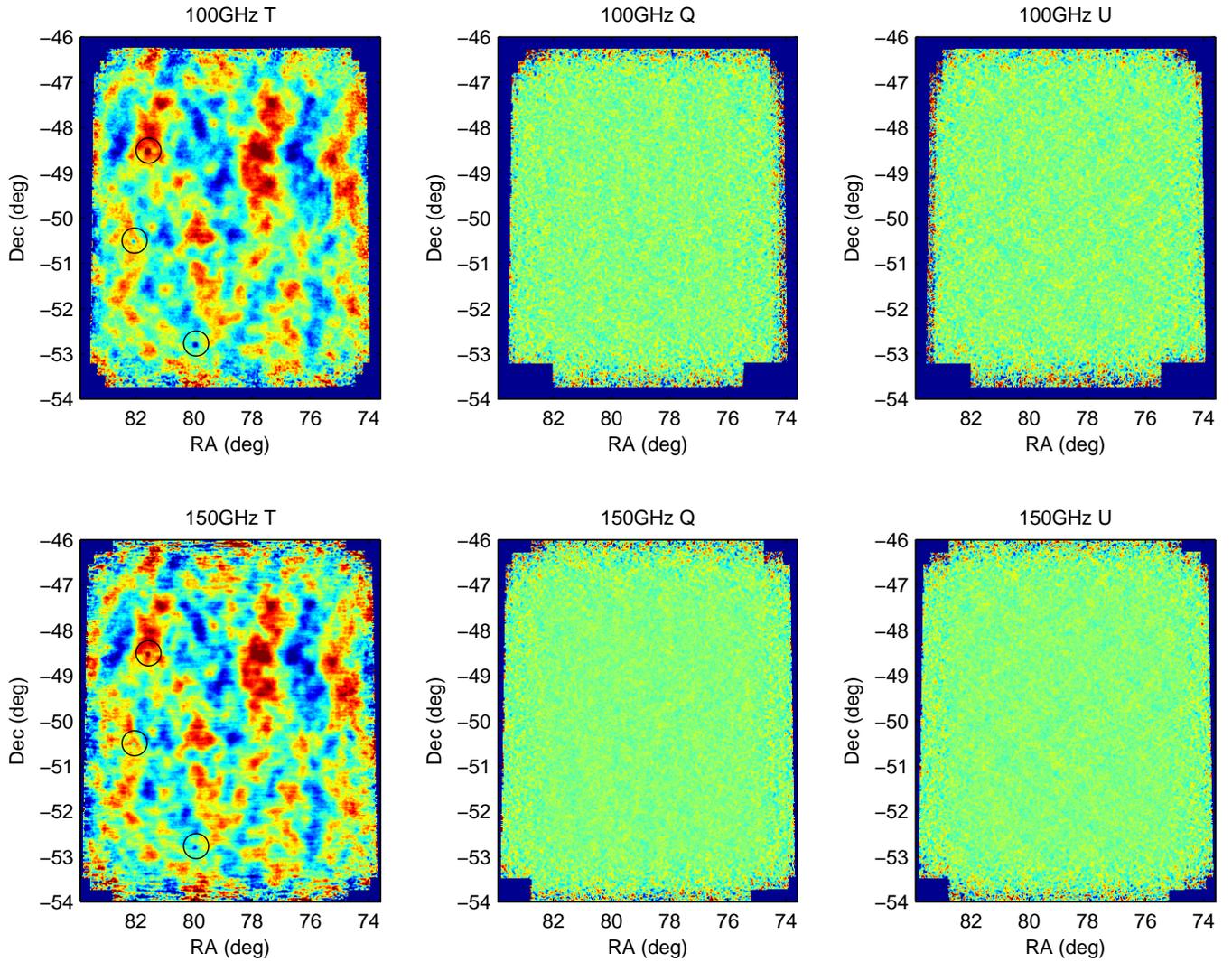}}
\caption{QUaD 100 and 150~GHz $T$, $Q$ and $U$ field difference maps.
The color scale is $\pm200$~$\mu$K in all cases, and the maps
have been smoothed with a $5'$ FWHM Gaussian kernel.
(Black circles indicate discrete sources which are
removed in the power spectrum analysis.)}
\label{fig:sigmaps}
\end{figure*}

Figure~\ref{fig:sigmaps} shows the 100 and 150~GHz $T$, $Q$ and $U$ maps
as generated using the process described in Section~\ref{sec:mapmaking}.
The signal to noise in the $T$ maps is extremely high as
is evident from the excellent spacial correlation of the
pattern between the two frequencies.
We deliberately plot the $Q$ and $U$ maps on the same color
scale to visually emphasize how small the polarization of the CMB
is compared to the degree scale structure in $T$.
Three discrete sources are weakly detected in the temperature maps,
but are not detected in polarization.

\subsection{Jackknife Maps}

The CMB polarization signal is extremely small necessitating
extreme attention to detail.
To probe for the presence of contaminating signal that
does not originate on the sky we perform a set of
jackknife tests.
For each test we split the timestream data into two approximately
equal subsets which should contain (nearly) the same sky signal, but
which might contain different false signal.
We then generate maps for each data subset
and difference them to produce jackknife maps:

\begin{eqnarray}
M^J & = & (M^1 - M^2)/2 \\
V^J & = & (V^1 + V^2)/4
\end{eqnarray}

\noindent where $M^J$ is the jackknife map, $M^1$ and $M^2$ are
the split maps and $V$ are the corresponding variance maps.
Note that we form jackknife maps with this normalization
by analogy to the non-jackknife map which is effectively
$(M^1 + M^2)/2$.

Plots of the jackknife maps are not presented here as they are
not particularly informative --- they look like noise.
However Section~\ref{sec:2dfpc} includes plots of the Fourier
transform of some sample jackknife maps,
and Section~\ref{sec:jackknifes} probes for any hint of structure
in these maps which is inconsistent with noise.

\section{Generation of Simulated Timestream}
\label{sec:sims}

Power spectra of maps such as those shown in Figure~\ref{fig:sigmaps}
need to be corrected for two effects: the ``noise bias'' (principally
affecting auto spectra), and the suppression of power due to the effects
of timestream filtering and finite beam size.
In addition we need to estimate the size of the final bandpower
uncertainties due to sample and noise variance.
We do this broadly following the MASTER technique~\citep{hivon02},
which requires accurate timestream level simulations of signal and noise.

\subsection{Signal simulations}
\label{sec:sigsims}

To construct simulations of signal timestream we start with LCDM
power spectra generated using the CMBFAST
program~\citep{zaldarriaga00} using the WMAP3 cosmological parameters given under the heading
``Three Year Mean'' in Table 2 of~\cite{spergel06},
and feed these into the ``synfast'' generator
(part of the HEALPix package\footnote
{See http://healpix.jpl.nasa.gov/index.shtml and~\cite{gorski05}})
to yield curved sky maps of $T$, $Q$ and $U$ at a resolution of
$0.4'$ (NSIDE of 8192).

We then read each day of data in turn, and loop over detectors.
For each we calculate the sky map $M_d$ which would be seen by
a detector of the given angle and polarization efficiency
by combining the $T$, $Q$ and $U$
input maps according to

\begin{equation}
M_d = \frac{1}{2} \left( M_T + \frac{1-\epsilon}{1+\epsilon} \left( M_Q \cos 2 \theta + M_U \sin 2 \theta \right) \right)
\end{equation}

\noindent where $\epsilon$ is the cross polar leakage
and $\theta$ is the detector polarization angle.
This ideal sky map is then convolved with an elliptical Gaussian
smoothing kernel with the appropriate parameters to simulate
the effect of the beam.
Finally we interpolate off the smoothed map along the pointing
trajectory for the given detector, computed as the observed
telescope pointing direction plus the detector
offset angle.

In Section~\ref{sec:polpar} we mentioned that
fixed nominal values are used for the detector
polarization angles and efficiencies when constructing
maps.
For each simulation realization we generate values
for each detector normally distributed about
these nominal values with the measured rms scatter
(1$^\circ$ and 0.015 respectively).
The simulated data is then re-mapped assuming the nominal
values as usual.

To simulate the small pointing wander mentioned in
Section~\ref{sec:pointing} we generate Gaussian
random numbers with zero mean and $\sigma$ of $0.5'$
as the RA and Dec offsets at the start and end of each
eight hour block.
We then linearly interpolate these offsets to each time-step
and add them in to the observed pointing trajectory.
Although in reality the wander has a more complex behavior
given the observation strategy, and averaged over hundreds of
days of data, this will lead to a broadening of the effective
beam width of the correct (very small) amount.

The detector offset angles are taken as fixed at our best estimate
values --- we have no evidence for time variation as mentioned
in Section~\ref{sec:detoffang} above.
Note that the measured beam centroid positions show repeatable
offsets between the two halves of each detector pair
with rms magnitude of $\sim 0.1'$.
When we construct maps using pair sum and difference data
we use the mean for each pair.
However when we sample from the sky to generate simulated
timestream we use the measured individual detector values.
Hence any $T$ to pol.\ mixing which occurs due to beam centroid mismatch
is fully included in the simulations --- and found to
be negligible --- see Section~\ref{sec:beammismatch}.

Note that generation of simulated signal timestreams takes
place before the field differencing operation --- effects
such as actual signal correlations and mismatch of pointing coordinates between lead 
and trail fields will therefore be included.

\subsubsection{Measurement of the beam widths}
\label{sec:beamwid}

To convolve the ideal sky map with the beam we need to know
the beam shape for each detector.
We have obtained this from nine days of special observations
of the bright quasar PKS0537-441, which is a point source at
the angular resolution of our experiment~\citep{fey00}.
The mean of the measured major and minor FWHM's is $5.0'$ at 100~GHz
and $3.5'$ at 150~GHz with uncertainties of $\approx 2$\%.
There is evidence for a small degree of variation in width between
the detectors of a given frequency band, and for $\leq 10$\% elongation 
(mismatch between the major and minor widths),
but for any given detector these are sub-dominant to the measurement uncertainty.
As when determining the detector offset angles from measurements of
RCW38 (see Section~\ref{sec:detoffang} above) we
have performed the elliptical Gaussian fits in both the timestream
and in maps yielding nearly identical beam widths and angles.
For further details see the Instrument Paper and our
Optics Paper~\citep{osullivan08}.

The effective beam width in our CMB field co-added maps is measured
with low signal to noise by the quasar PKS0524-485
(the brightest of the circled sources in Figure~\ref{fig:sigmaps}). 
This source appears as a Gaussian peak with width consistent
with the single day observations of PKS0537-441.

\subsection{Noise simulations}
\label{sec:noisims}

As shown in Figures~\ref{fig:timestream} and~\ref{fig:todspec}
the detector timestream is dominated by heavily correlated
low frequency atmospheric noise.
The goal of the timestream noise generator is to reproduce
simulated half-scans which are indistinguishable from the
real under a battery of tests in both the time
and frequency domains.

To achieve this we find it necessary to measure and re-generate
longer pieces of timestream, and then cut them down
to the half-scans which are actually used when constructing
the maps.
First the complete time period spanning each five scan set 
is Fourier transformed (including turn arounds).
Then for each of a set of logarithmically spaced frequency bins
we take the covariance matrix
of the Fourier modes between all channels.
This matrix is Cholesky decomposed and used to mix
uncorrelated random numbers to re-generate the observed
degree of covariance.
This process is repeated for each frequency bin
and then the resulting sets of Fourier modes are inverse
transformed to yield simulated timestream.
Possible correlations between the real and imaginary
parts of each Fourier mode are preserved by using
complex covariance and Cholesky matrices, but
this process assumes the Fourier modes are correlated
only between channels, and not between modes.
Breaking the simulated timestream down into half-scans
reintroduces such correlations and
we find it has equivalent temporal and spectral
characteristics to the real data under a variety
of statistical tests.

As an example the right part of Figure~\ref{fig:todspec}
shows the power spectra
of the resulting simulated timestream --- comparing to
the real PSDs in the left part of the figure
we see an excellent match.
Note that the simulation produces non pair differenced
timestream which has then been differenced to generate
the spectra shown in Figure~\ref{fig:todspec} --- it is clear that the
channel-channel correlations are being reproduced to high accuracy.
The narrow lines in the real spectra are re-generated as broader
boxcar features due to the frequency bin width used in the
simulation process --- this is irrelevant since these
frequencies are far above the science band.

The signal to noise ratio in the timestream data is
sufficiently low that we do not need to
subtract the signal contribution before taking the noise
spectra.
(The PSD of signal only LCDM timestream
is more than 2 order of magnitude below the noise
at all frequencies for the pair sum, and 4 orders of
magnitude for the pair difference.)
To confirm that the noise modeling operates correctly we
have performed a ``sim the sim'' study where the output
of a single signal plus noise timestream simulation is used as
the input to a complete simulation set, passing all the jackknife
tests described in Section~\ref{sec:jackknifes} as expected.

\subsection{Simulated maps}

We generate many realizations of signal and noise timestream
and produce the full set of un-split and jackknife split maps from
these using the same code used to make maps of the real
timestream in Section~\ref{sec:mapmaking} above.
For each realization we then add the signal and noise maps
to form signal plus noise maps --- since the map making
process is linear this is equivalent to adding the timestreams,
and computationally more efficient.
The generation of simulation realizations is computationally
costly and hence their number is relatively small (500 in this
analysis).

\section{From Maps to Power Spectra}
\label{sec:maps2spec}

This section will describe the various steps which take us from
the real and simulated maps to angular power spectra.

\subsection{The apodization mask}
\label{sec:apmask}

In addition to the signal maps $M_{T,Q,U}$ the process described in
Section~\ref{sec:mapmaking} also produces maps of
estimated variance $V_{T,Q,U}$ based on the variance of each co-added half-scan,
assuming that the noise is white.
Empirically these are found to be close to correct for
the polarization maps and a 10\% (20\%) underestimate
for the 100~GHz (150~GHz) $T$ maps, as expected since
the pair sum data contains significant $1/f$ noise.
We take the inverse of these variance maps as the apodization
masks:

\begin{equation}
A_X = \frac{1}{V_X}
\end{equation}

However due to the partially overlapping ``tiles'' of coverage
resulting from our observation strategy and the differing detector
offset angles there are sharp steps in the masks (of $\sim10$\% magnitude).
If these steps are not smoothed out then the product
of the map and mask (the quantity which is about to get
Fourier transformed) will also have sharp steps.
Such steps correspond to a mixing of power from large to small
scales which is particularly undesirable in the $TT$ spectra
where even a small fractional contribution from the
very large low multipole $C_\ell$ values would dominate
over the intrinsic power in the damping tail region
of the spectrum\footnote{
Note that the relevant ratio is in $C_\ell$, not $\ell(\ell+1)C_\ell$
which is the quantity conventionally plotted ---
for the $TT$ spectrum under LCDM $C_{200}/C_{2000}\approx 2500$}.
To mitigate this effect we convolve the inverse variance mask
with a Gaussian shaped smoothing kernel of FWHM 0.5$^\circ$:

\begin{equation}
A'_X = A_X \ast G
\end{equation}

We emphasize that an arbitrary apodization mask can be used without
biasing the results and, in fact, the inverse variance mask
is only optimal in the limit of low signal to noise which is
not the case for the $T$ maps.
In principle an optimal mask could be generated for each spectrum,
and in fact each bandpower within each spectrum, based on its
signal to noise, but we have not pursued this complication
in this analysis.

After smoothing we inject Gaussian shaped ``divots'' with FWHM of 0.5$^\circ$
into the apodization masks at the locations of three discrete
sources circled in Figure~\ref{fig:sigmaps} to null out any
effect they might otherwise have on the results:

\begin{equation}
A''_x = A'_X \left( 1 - \sum_i{ \exp \left( - \left( \frac{(x-x_{o,i})^2 +
                        (y-y_{o,i})^2}{\sigma^2} \right) \right)} \right)
\end{equation}

\noindent where $x$ and $y$ are the pixel coordinates, $x_{o,i}$
and $y_{o,i}$ are the source locations, and the loop $i$
is over the set of sources to be nulled.
Note that although the sources are visible only
in the $T$ maps we apply the same masking in
polarization also.

The final apodization masks imply an effective sky area
for this analysis of $\approx 25$~square degrees.

\subsection{Fourier Transform and Power Spectra}
\label{sec:ft_and_ps}

We next make the flat sky approximations
and take the two dimensional discrete Fourier transform of the product
of the map and apodization mask:

\begin{equation}
m_X = c \ \mathrm{FT}(M_X A''_X)
\label{do_ft}
\end{equation}

\noindent where $c$ is a normalization constant which acts to
make the apodization and Fourier transform operations total power
preserving.

In the Fourier domain the transform from $Q,U$ to $E,B$
is simply

\begin{eqnarray}
  m_E & = & +m_Q \cos 2 \phi + m_U \sin 2 \phi \\
  m_B & = & -m_Q \sin 2 \phi + m_U \cos 2 \phi
\end{eqnarray}

\noindent where $\phi$ is the polar angle of each mode $m$
with respect to the Fourier plane origin.

We then take products of the (complex) modes
within and between each set

\begin{equation}
p_{XY} = m_X m^*_Y
\end{equation}

\noindent where $X$ and $Y$ can be $T$, $Q$ or $U$ at either
100 or 150~GHz.

Each set of mode products $p_{XY}$ is then multiplied by $d = \ell(\ell+1)/2\pi$
where the multipole $\ell = 2\pi u$, and $u$ is the Fourier
conjugate variable to angular distance from the map center.
Finally we take the binned angular power spectra

\begin{equation}
b_{XY,i} = \left\langle d p_{XY,i} \right\rangle
\label{eqn:mk_bandpowers}
\end{equation}

\noindent where the mean is taken over the modes
in each annulus $i$ of the Fourier plane.
The multiplication of the map by the mask in real
space corresponds to a convolution in Fourier space ---
hence the Fourier modes $m_X$ are correlated and so are
the bandpower estimates $b_{XY}$.
For this analysis we set the bin spacing to $\Delta \ell =81$
which results in correlation of $\sim20$\%.
No importance should be attached to the exact value 81 ---
it was an arbitrary choice made early on and remained
unchanged throughout the development of the analysis.

The above description is a slight simplification of the
actual procedure used.
In reality for each cross spectrum $XY$ we enforce a common
apodization mask $A''_{XY} = \sqrt{A''_X A''_Y}$ before taking
the Fourier transforms and proceeding to power spectra.

\subsection{Fourier Plane Masking}
\label{sec:2dfpc}

The strong theoretical expectation is that the CMB
is isotropic on the sky and therefore also in the two
dimensional Fourier plane.
The noise however is certainly not --- atmospheric variation
injects strong low frequency fluctuations ($1/f$ noise) along
the scan direction and hence in a band around the perpendicular
direction in the Fourier domain.
We are perfectly at liberty to down weight or excise portions
of the Fourier plane when we take power spectra --- no bias
will result so long as the weight/cut criteria are independent
of the mode values of the real maps.

\begin{figure}[h]
\resizebox{\columnwidth}{!}{\includegraphics{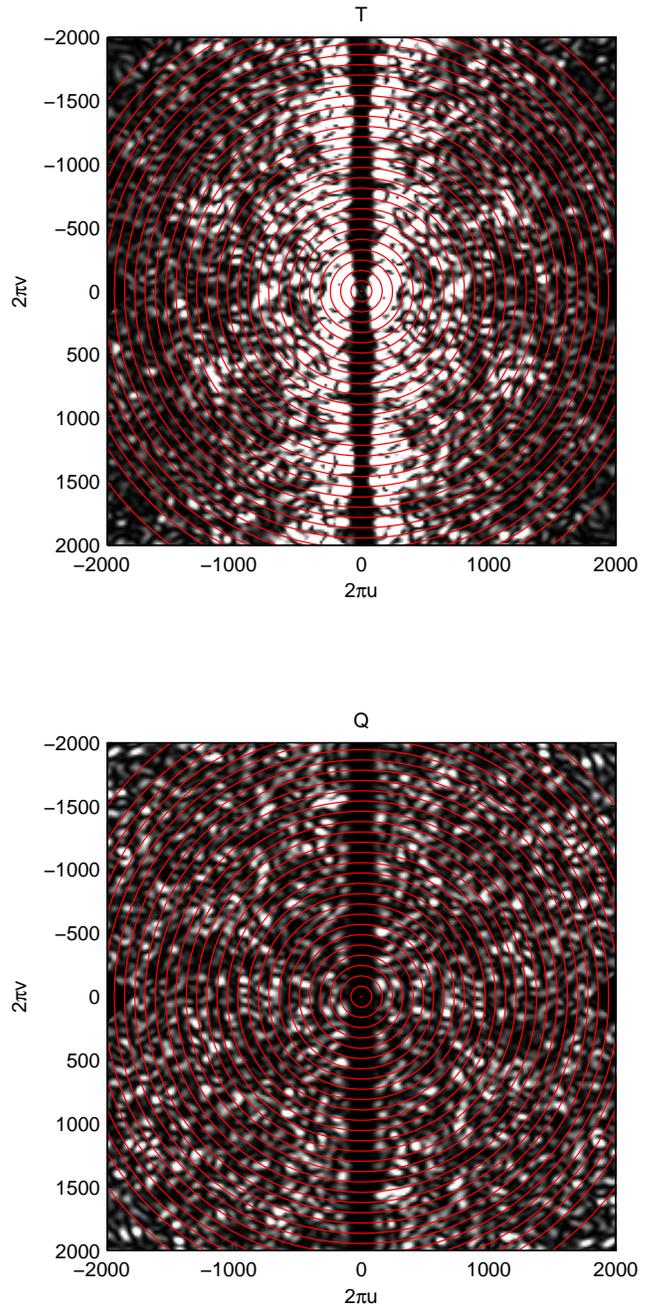}}
\caption{Fourier transforms of the apodized 150~GHz $T$ and $Q$ deck
jackknife maps.
The real part of the square of the Fourier modes is shown with
a linear color stretch.
The red circles indicate the annuli within which means are taken
to generate bandpower values.
Note that the horizontal direction in this plot corresponds to
the scan direction --- see text for further details.}
\label{fig:2dfp}
\end{figure}

Figure~\ref{fig:2dfp} shows the 150~GHz $T$ and $Q$ Fourier
modes for the deck jackknife map.
The third order polynomial filter which has been applied
to each half-scan removes power along the scan direction
producing the dark vertical bands down the middle of the plots.
Atmospheric $1/f$ which survives the filtering appears
as a brighter band either side of the dark band in the $T$
plot, but due to the unpolarized nature of the atmosphere, does
not appear in the $Q$ plot (see Section~\ref{sec:noisims}
above).
Because the atmospheric noise is highly correlated amongst
adjacent detector pairs in the array we in fact see a complex
structured pattern of non-uniform noise in the $T$ plot ---
our noise simulations fully reproduce this pattern.

Section~\ref{sec:apmask} mentioned the need to smooth
the inverse variance derived apodization mask to reduce
up-mixing of power from small to large multipoles.
There is an additional similar, but smaller, effect.
Due to the detector offset angles each has observed a
slightly different rectangular patch of the CMB sky.
We polynomial filter each detector's timestream for each
half-scan to reduce atmospheric noise, and so
the CMB modes removed from each ``tile'' are slightly
different.
Hence when we co-add the data from all detectors
small step artifacts are present in the overall map
at the tile overlap boundaries.
Examining the non-jackknife Fourier $T$ modes using plots
similar to those shown in Figure~\ref{fig:2dfp} we see
up-mixed power becoming dominant over the intrinsic for
multipoles at $\ell>1800$ in a narrow band around the
scan direction axis --- as expected, since the artifacts
are vertical step edges in the maps.
This spurious power is also seen in the signal only
simulations confirming its origin is up-mixing.
Note that this effect is only significant in $T$
due to the much larger ratio between low and high
multipoles than in the polarization spectra.
However we excise these modes when generating all power
spectra resulting in a trivial increase in the uncertainty
of our highest bandpowers ($<3$\%).
This effect will be more important in a future analysis
extending to $\ell=3000$.

One could weight the Fourier modes when combining
to form power spectra based on their signal to noise ratio
as measured in the signal and noise simulations.
This would be (at least a partial) substitute for the polynomial
half-scan filtering --- one either removes the atmospheric
$1/f$ in the timestream or in the two dimensional Fourier
plane.
Looking at Figure~\ref{fig:2dfp} such Fourier plane weighting
would clearly be beneficial for the $TT$ spectra.
Although we have experimented with such weights and cuts
we have not implemented them in the analysis presented
here --- the benefit for the essentially white noise
polarization spectra is close to zero.

\subsection{Raw spectra compared to simulations}

\begin{figure}[h]
\resizebox{\columnwidth}{!}{\includegraphics{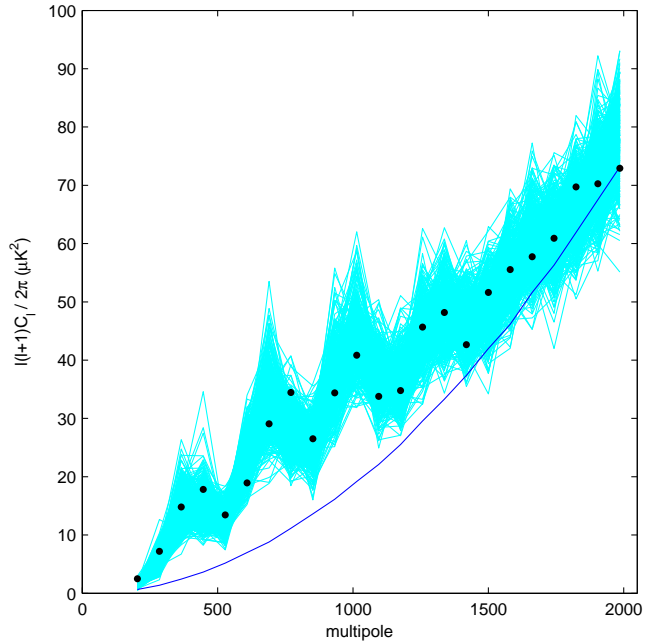}}
\caption{Comparison of real and simulated raw 150~GHz $EE$ spectra.
The cyan curves are the ensemble of signal plus noise simulations,
the black points are the observed bandpower values, and the blue
curve is the mean of the ensemble of noise only simulations.}
\label{fig:rawrealsimspec}
\end{figure}

By taking the mean of the masked Fourier mode products within
each annulus we generate raw power spectra of the real and simulated maps.
Figure~\ref{fig:rawrealsimspec}
compares the 150~GHz $EE$ spectra and is in a sense the fundamental
result of our analysis --- is the observed spectrum consistent
with being a realization of LCDM plus noise?
To test that hypothesis we could simply perform $\chi^2$ tests at this point.
However, for presentation purposes, and to allow our spectra
to be used for cosmological parameter analyses, we proceed
to noise and filter correct them as follows.

\subsection{Removing the noise bias}

At this point in the analysis we have sets of bandpower
values calculated using Equation~\ref{eqn:mk_bandpowers}
for each of the auto and cross spectra.
We have these for the real spectra, and each realization
of the signal only, noise only, and signal plus noise
simulations.
We will denote these $\mathbf{r}_{XY}$, $\mathbf{s}_{XY}$,
$\mathbf{n}_{XY}$ and $\mathbf{sn}_{XY}$ respectively.

To remove the noise bias we take the mean of the ensemble
of noise only simulations and subtract from the
real spectra

\begin{equation}
\mathbf{r}'_{XY} = \mathbf{r}_{XY} - \overline{\mathbf{n}_{XY}}
\label{eqn:noise_debias}
\end{equation}

\noindent
In Figure~\ref{fig:rawrealsimspec}
the blue curve is subtracted from the black points.
Correlated noise, presumably from the atmosphere,
is fully simulated by the process described in
Section~\ref{sec:noisims} above.
Such noise can result in the mean noise cross spectra
--- where cross spectrum means either within, or across
frequency bands --- being non zero.
In practice the levels are very small compared to the auto
spectra but for completeness we subtract them anyway.

\subsection{Band power window functions}
\label{sec:bpwf}

As mentioned in Section~\ref{sec:ft_and_ps}
multiplication by the apodization mask in image space
corresponds to convolution in the Fourier plane by
the Fourier transform of that mask.
In Figure~\ref{fig:2dfp} we see that adjacent Fourier modes
(pixels) are highly correlated, and that the resulting annular
mean bandpower estimates will therefore also be correlated.
We thus need to know how much each multipole on the sky
contributes to each experimental bandpower --- the so called ``band power
window functions'' (BPWF's)~\citep{knox99}.
This could be determined by running sets of signal
only simulations, each with non-zero input power only in a narrow band
of multipoles.
However the computational cost of doing this in practice is too high,
and we hence use an alternate, much faster method.

\begin{figure}[h]
\resizebox{\columnwidth}{!}{\includegraphics{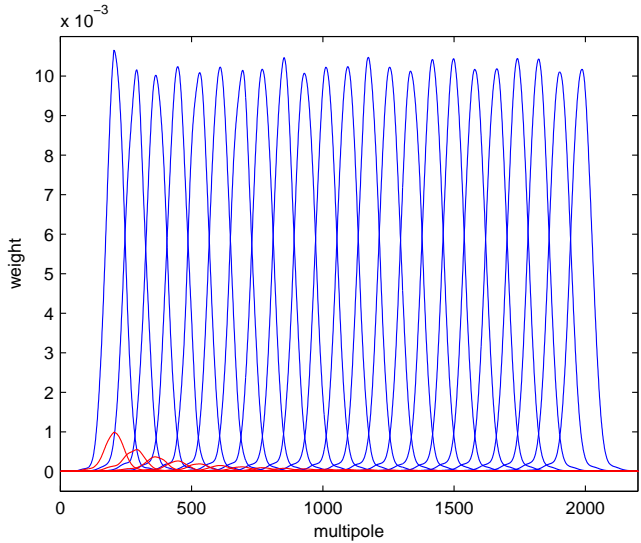}}
\caption{The bandpower window functions for the 150~GHz
$BB$ spectrum.
Each blue curve shows the relative response to $BB$ sky
power, while the corresponding red curves show the response
to $EE$.}
\label{fig:bpwf}
\end{figure}

We take each narrow annulus in the Fourier plane in turn
and convolve this with the Fourier transform of the mask.
For $Q$ and $U$ we generate annuli corresponding to pure
$E$ or $B$ modes.
Taking the power spectrum of the convolved annuli measures
the response of each experimental bandpower to sky power
in the given annulus.
Placing these spectra into the rows of a matrix, the
columns are then the desired BPWF's (interpolating to
every multipole).
For $BB$ we have two functions per bandpower representing
the response to true $BB$ sky power, and also the response to $EE$.
In practice the mixing is $\sim10$\% in our lowest bandpower
and falls rapidly with increasing $\ell$ (see Section~\ref{sec:ebmix}).
Figure~\ref{fig:bpwf} shows some example BPWF's.

\subsection{Determining the filter/beam suppression factor}
\label{sec:filtbeamcorr}

Using the BPWF's described above
we can calculate the expectation values $\mathbf{e}_{XY}$
of the observed bandpowers given the input LCDM power spectra used to
generate the signal simulations.
We then take the ratio of the mean of the signal only simulated spectra
to these expectation values to empirically
determine the suppression of power which has occurred
due to the convolution of the sky by the telescope
beam pattern, and the polynomial filtering of the timestream

\begin{equation}
\mathbf{f}_{XY} = \frac{\overline{\mathbf{s}_{XY}}}{\mathbf{e}_{XY}}
\end{equation}

\begin{figure}[h]
\resizebox{\columnwidth}{!}{\includegraphics{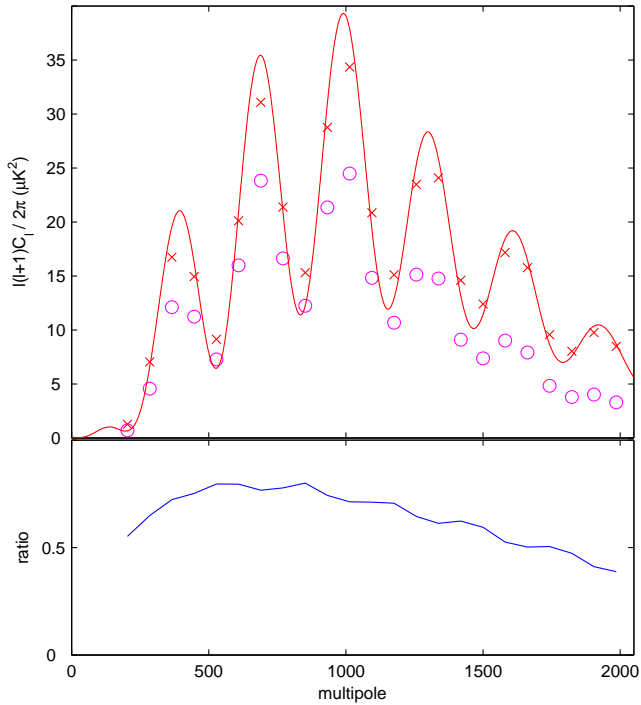}}
\caption{{\it Upper:} Input $EE$ spectrum (red line), calculated
expectation values (red crosses), and mean simulated signal only
spectrum (magenta circles).
{\it Lower:} The ratio of the two sets of points in the upper panel --- 
the filter/beam suppression factor.
(This plot is for the 150~GHz spectrum.)}
\label{fig:filtbeam}
\end{figure}

\noindent --- this is the factor by which the real spectra must be divided
to yield an un-biased estimate of the true sky power.
Note that this process will automatically include any simulated suppression
effect (including the so-called "pixel window functions").
Figure~\ref{fig:filtbeam} illustrates this step --- the curve
in the lower panel approaches the ``beam window function''
$W_\ell$ at higher $\ell$.

We now divide the spectra by their respective suppression factors

\begin{equation}
\mathbf{r}''_{XY} = \frac{\mathbf{r}'_{XY}}{\mathbf{f}_{XY}}
\label{eqn:filterbeam_corr}
\end{equation}

\noindent In practice we apply the $\mathbf{f}_{TT}$ correction
to the $TT$ spectra and the $\mathbf{f}_{EE}$ correction to the
$EE$, $BB$ and $EB$ spectra.
For the $TE$, $TB$ spectra we use the geometric mean
of $\mathbf{f}_{TT}$ and $\mathbf{f}_{EE}$.

\subsection{Power spectrum results}

We also apply the noise de-bias and filter/beam correction
operations of Equations~\ref{eqn:noise_debias} and~\ref{eqn:filterbeam_corr}
to each signal plus noise realization.
Their fluctuation then provides an estimate
of the uncertainty of the real bandpower values.
This uncertainty will only be correct in as much as the
theory spectrum used as input to the signal simulations
matches reality --- if there is any significant disagreement
one should iterate the entire process until it converges.
As we will see in Section~\ref{sec:comp2lcdm} our results are
in fact perfectly compatible with the WMAP3 based model used
as input to the signal simulations (see Section~\ref{sec:sigsims})
so there is no need to iterate.

\begin{figure*}[h]
\resizebox{\textwidth}{!}{\includegraphics{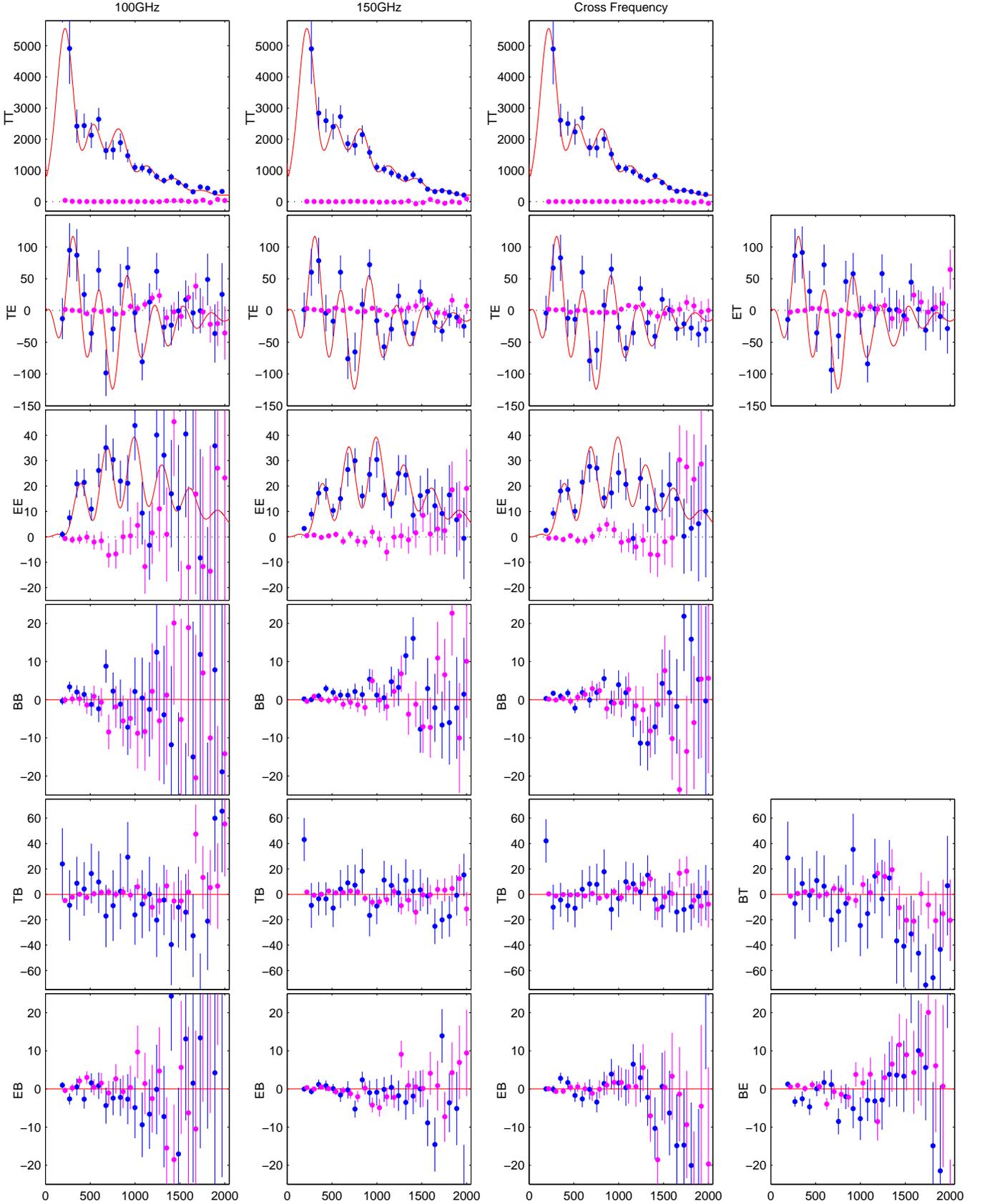}}
\caption{The full set of QUaD power spectrum results.
The blue points are the signal spectra, the magenta the deck jackknife
(see text for definition), and the red curves a conventional LCDM model.
The horizontal and vertical axes are multipole $\ell$ and $\ell(\ell+1)C_l/2\pi$
($\mu$K$^2$) respectively.
Each bandpower is $\sim20$\% correlated with its neighbors and, where
signal dominated, there are strong correlations between bandpowers across
frequency bands.
The error bars are the standard deviation of the simulated signal
plus noise bandpower values, and the two sets of points have been
offset by $\pm15$ in $\ell$ from their nominal values for clarity.
Note that absolute calibration and beam size uncertainty are not
included in the error bars.
The jackknife spectra are consistent with null --- see Table~\ref{tab:ptes}
and the text.}
\label{fig:spec_res}
\end{figure*}

Figure~\ref{fig:spec_res} shows the full set of 21 signal spectra:
within each frequency band we have $TT$, $TE$, $EE$, $BB$, $TB$
and $EB$, while across frequency bands we can
form $T_{100}T_{150}$, $T_{100}E_{150}$, $E_{100}T_{150}$,
$E_{100}E_{150}$, $B_{100}B_{150}$, $T_{100}B_{150}$, $B_{100}T_{150}$,
$E_{100}B_{150}$ and $B_{100}E_{150}$.
We see that the 150~GHz spectra have better sensitivity than the
100~GHz --- this is partly due to the larger number of detectors
at the higher frequency, but mostly due to the smaller beam size
(and hence smaller beam correction).

\subsection{Alternate analysis}

\begin{figure}[h]
\resizebox{\columnwidth}{!}{\includegraphics{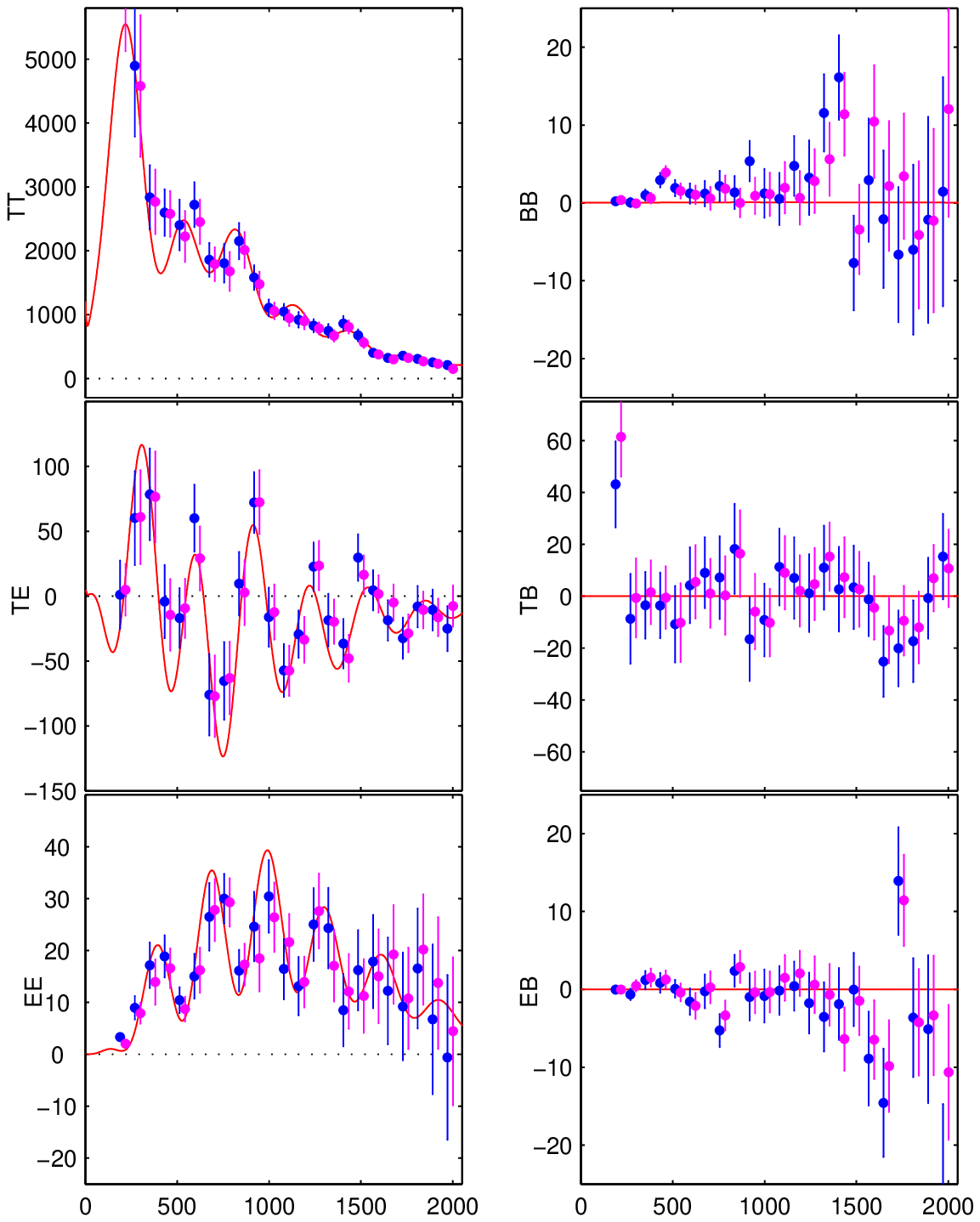}}
\caption{Comparison of 150~GHz spectra for the main (blue points)
and alternate (magenta points) pipelines.
The horizontal and vertical axes are multipole $\ell$ and $\ell(\ell+1)C_l/2\pi$
($\mu$K$^2$) respectively,
and the two sets of points have been
offset by $\pm15$ in $\ell$ from their nominal values for clarity.}
\label{fig:2anal}
\end{figure}

The main analysis in this paper comes from an evolution
of ``pipeline 2'' in our previous paper~\cite{ade07}.
A second pipeline exists, the
low level parts of which are derived from our
previous ``pipeline 1'', but which, for the results
presented here, is now also performing flat sky power spectrum
estimation.
This pipeline currently has noise modeling
that is somewhat less sophisticated than that described in
Section~\ref{sec:noisims}.
The two pipelines were independently written and share no code.
Although the algorithms implemented are intended to be basically
the same they are sure to differ in some (hopefully unimportant)
details.
The fact that the final results agree very closely therefore
adds considerable additional confidence.
Figure~\ref{fig:2anal} compares the 150~GHz spectra from
the two pipeline.

\section{Jackknife tests}
\label{sec:jackknifes}

To probe the results shown in Figure~\ref{fig:spec_res} for systematic
contamination we conduct a battery of jackknife tests.
As mentioned above split dataset maps of the real data and
each simulation realization are generated.
Each split divides the timestream data into two approximately
equal subsets which should contain (nearly) identical sky signal,
but which might contain different contaminating signal.
In this analysis we consider four such data splits which we call
the deck angle jackknife, scan direction jackknife, split season
jackknife and focal plane jackknife.
Each of these is described in more detail below.

For each jackknife the split maps are subtracted from one another,
divided by two, and we then proceed to power spectrum estimation as usual.
For the real spectra
we then calculate the $\chi^2$ versus the null model

\begin{equation}
\chi^2 = \mathbf{r}'' \mathbf{C}^{-1} \mathbf{r}''^t
\label{eqn:chisq}
\end{equation}

\noindent where $\mathbf{C}$ is the bandpower covariance matrix as estimated from the ensemble
of signal plus noise simulations.
We also calculate $\chi^2$ for each signal plus noise simulation.
The generation of simulation realizations is computationally
costly and hence their number is relatively small (500 in this
analysis).
This has two implications.
Firstly, since bandpower correlations beyond nearest neighbor
are sufficiently weak as to be lost in the measurement noise, we 
set all but the main and first two off diagonals of the
bandpower covariance matrix to zero.
Secondly when calculating the $\chi^2$ values for the signal plus noise
realizations extreme fluctuations have the opportunity to
partially self-compensate by injecting extra covariance
into the matrix which they will be measured against, biasing the
resulting $\chi^2$ values low.
To get around this we re-calculate the covariance
matrix excluding each realization in turn before calculating
the $\chi^2$ for that realization.

\begin{figure*}[h]
\resizebox{\textwidth}{!}{\includegraphics{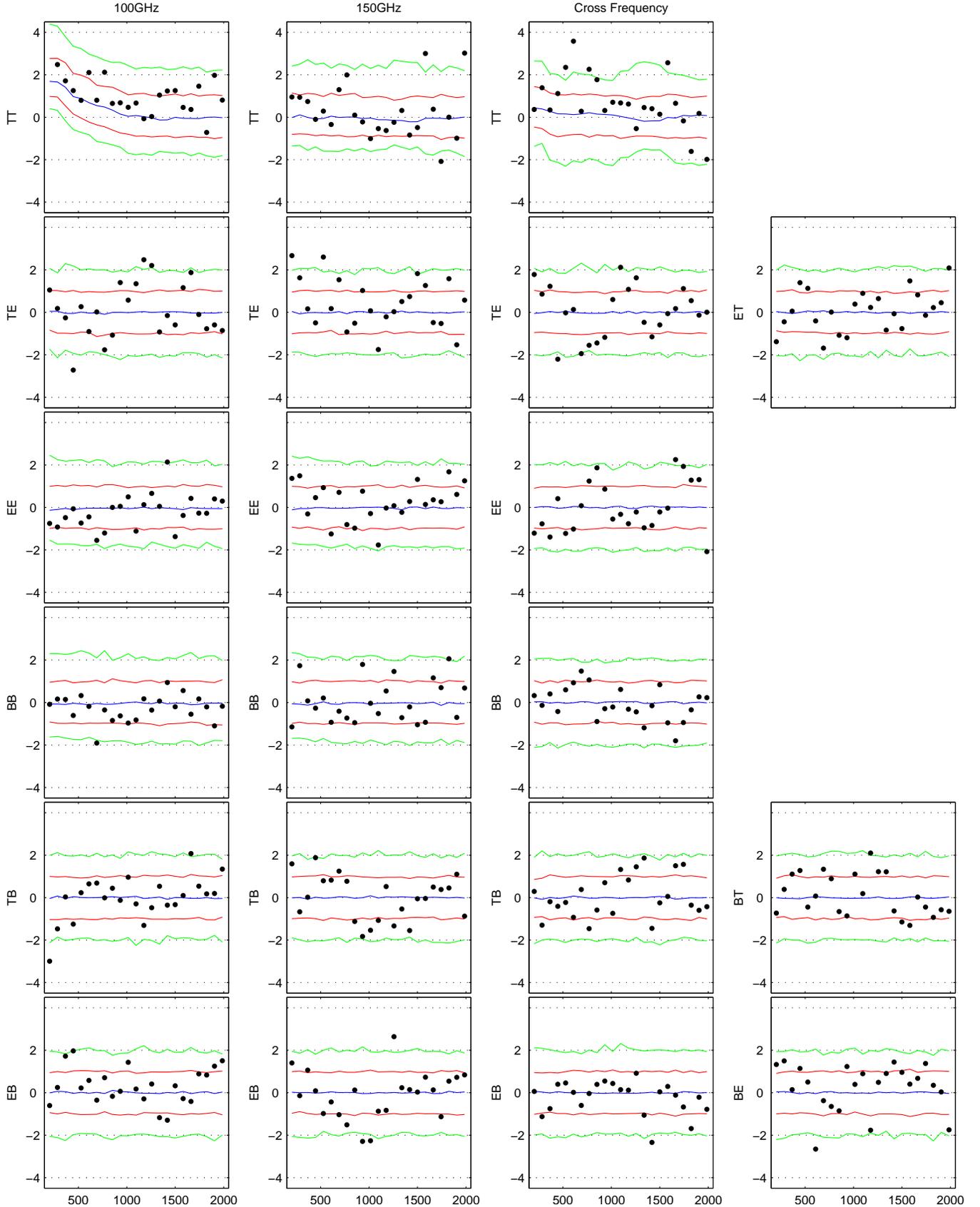}}
\caption{Bandpower deviations of the deck jackknife spectra.
The horizontal axis is multipole number.
The spectra shown as magenta points in Figure~\ref{fig:spec_res} have
been divided by their error bars to investigate the
contribution of each bandpower to $\chi^2$.
The green, red and blue lines show the 2.3\%, 15.9\%, 50\%, 84.1\% and 97.7\%
points of the integral distribution of the simulated signal plus noise
bandpower values.
}
\label{fig:bpdevs}
\end{figure*}

We find that the simulated jackknife $\chi^2$ distributions
for $TT$ do not follow the analytical expectation.
Examining Figure~\ref{fig:bpdevs}, which shows the bandpower
deviations of the deck jack spectra, we see that there are
two reasons why.
Firstly the simulated bandpower distributions
are significantly non-Gaussian --- the colored lines do not sit
at $-2$ through $+2$.
Secondly, and more importantly, there is significant
predicted imperfect cancellation --- the simulated distribution
deviates strongly from a median of zero at large angular scales.
The reason for this latter effect was mentioned in
Section~\ref{sec:2dfpc} ---
in any split where the sky coverage ``tiling'' for the two
subsets is non-identical the interaction of the true
sky brightness distribution and the half-scan polynomial filtering generates
slightly different output maps.
Consistent with our overall simulation based approach we therefore
take the probability to exceed (PTE) the real spectra $\chi^2$
values versus the simulated distributions, rather than the
analytical.

\begin{figure}[hhhhh]
\begin{center}
\resizebox{0.7\columnwidth}{!}{\includegraphics{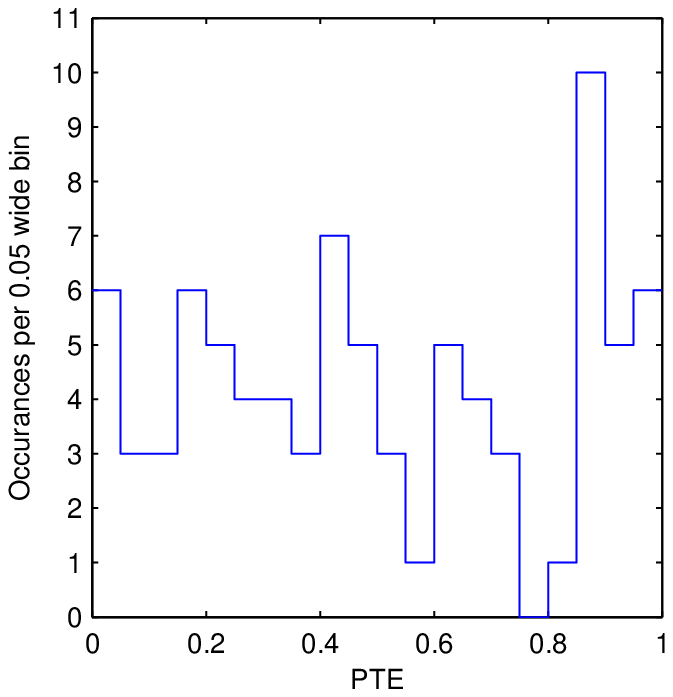}}
\end{center}
\caption{Distribution of the $\chi^2$ PTE values from Table~\ref{tab:ptes}.}
\label{fig:pte_dist}
\end{figure}

Table~\ref{tab:ptes} shows the full set of PTE values for all
the jackknifes and spectra --- there are no strong indications
of problems.
These values are expected to be uniformly distributed between
zero and one, and in Figure~\ref{fig:pte_dist} they are seen to be so.
See the following sub-sections for detailed discussion of these results.

\begin{deluxetable}{c c c c c}
\tablecaption{Jackknife PTE values from $\chi^2$ tests \label{tab:ptes}}
\tablehead{\colhead{Jackknife} &
\colhead{100~GHz} & \colhead{150~GHz} & \colhead{Cross} & \colhead{Alt.\ Cross}}
\startdata
\input{ptetab}
\enddata
\end{deluxetable}

\subsection{Deck jackknife}

The deck angle jackknife is perhaps the most powerful.
Due to the locking of our daily observations to LST
the two halves of this split contain data taken over
completely different ranges of telescope azimuth.
In addition, and equally importantly, the entire
telescope is rotated by 60$^\circ$ around the line of sight
between the two observation sessions.
The two data subsets are therefore separated from one another
in time, azimuth angle, and detector polarization
angle as projected on the sky.
It is very hard to conceive of a source of contamination
which would be common in the polarized maps for the two
halves of this split.
We see no significant indication of problems with the deck angle
jackknife in Figures~\ref{fig:spec_res} and~\ref{fig:bpdevs},
or in Table~\ref{tab:ptes}.
One might perhaps worry about the rather
low values for 150~GHz and cross-frequency $TT$ but we caution
against over-interpreting such numbers --- the table
contains 84 numbers so on average four numbers below 0.05
are to be expected.
In addition looking at Figure~\ref{fig:bpdevs} we see that
the bandpower making the strongest contribution to the cross
frequency $\chi^2$ has a very low deviation in the
150~GHz spectra.
Note that Figure~\ref{fig:pte_dist} shows no obvious excess of low
PTE values.

\subsection{Scan direction jackknife}

The scan direction jackknife splits the data into the
out-going and returning half of the scans.
In terms of external contamination this is perhaps the
easiest test to pass --- only something very rapidly varying
would cause it to fail.
However it is a stringent test for internal instrumental effects.
Any scan synchronous false signal, caused perhaps by
motion of the liquid cryogens in the tanks provoked by
the telescope motion, would likely fail to cancel in this
jackknife.
Also any failure to adequately deconvolve the temporal
response of the detector channels would cause this
test to fail.
Looking at Table~\ref{tab:ptes} we see no problems.

\subsection{Split season jackknife}

For the split season jackknife we have divided the time ordered
list of days in half (not in fact into 2006 and 2007 --- the split
occurs in late August 2006).
We then make maps with each set of days and difference
them.
This test would fail if there were a significant shift
in the absolute calibration of the telescope system
beyond the atmospheric loading effect whose correction is described
in Section~\ref{sec:gainsupcor}.
Looking at Table~\ref{tab:ptes} we see a few
numbers below $0.05$ but nothing highly significant.
Examining the bandpower deviations (a plot
not shown in this paper analogous to Figure~\ref{fig:bpdevs})
we find that for each spectrum with a low PTE it is caused
by a different bandpower(s), giving no further hint
of any problem.

\subsection{Focal plane jackknife}

This jackknife splits the detectors into the two
orientation groups which are separated by 45$^\circ$ --- referred
to as instrument-$Q$ and instrument-$U$.
Due to co-adding across the two deck angles each map
pixel has still been observed at two polarization angles
allowing the construction of $Q$ and $U$ maps as usual.
This test is perhaps the weakest, but might reveal
problems with instrumental false signal in a subset of
the pairs (although it is hard to see how that would not
also show up in the deck angle jackknife).
Looking at Table~\ref{tab:ptes} we see no problems.
Although this jackknife includes the lowest number
in the table (0.008) one such number is not improbable
and we note again that the PTE distribution
shown in Figure~\ref{fig:pte_dist} is consistent
with uniform.

\section{Foreground Studies}
\label{sec:foregrounds}

\subsection{Frequency difference maps and spectra}

We can also take the difference between our 100~GHz and 150~GHz maps.
It is important to be clear that this is not a jackknife in the
sense of Section~\ref{sec:jackknifes} --- the true sky brightness
distribution may in fact differ at these two frequencies.
Therefore any failure to cancel might be due to the presence
of astrophysical foregrounds, as well as instrumental systematics or
contamination.

\begin{figure}[h]
\begin{center}
\resizebox{0.65\columnwidth}{!}{\includegraphics{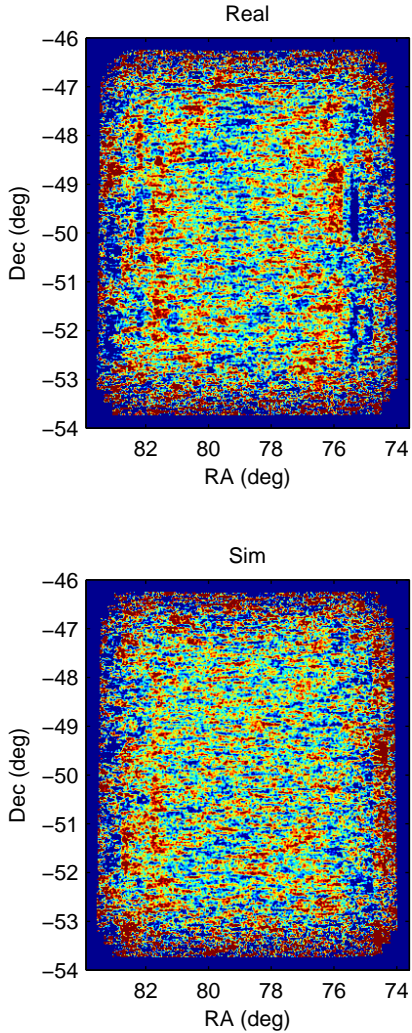}}
\end{center}
\caption{{\it Upper:} The difference between the 100 and 150~GHz $T$ maps
shown in Figure~\ref{fig:sigmaps} on a $\pm20$~$\mu$K color
stretch.
{\it Lower:} The same thing for a signal plus noise simulation
realization.}
\label{fig:fjackmaps}
\end{figure}

As described in Section~\ref{sec:abscal} we find the
absolute calibration scalings for our 100 and 150~GHz maps
by cross correlating them against the same B03 maps (which are at 150~GHz).
However, if the sky pattern differs at these two frequencies then 
this will still show up in the frequency difference maps.
Figure~\ref{fig:fjackmaps} shows the difference between the
real 100 and 150~GHz $T$ maps, and the same thing for
a signal plus noise simulation realization.

As mentioned in Section~\ref{sec:jackknifes}, when subtracting
maps with different sky coverage ``tiling'' a small degree of
mismatch is expected due to the polynomial filtering.
This is the cause of the vertical ``step edges'' observed in the upper
panel of Figure~\ref{fig:fjackmaps}.
The simulation realizations show similar effects, although in the
real map they do appear to be unusually strong.
Note that we also expect to see cancellation failure at smaller
angular scales due to the differing beam sizes for the two
frequency bands.

\begin{figure}[h]
\resizebox{\columnwidth}{!}{\includegraphics{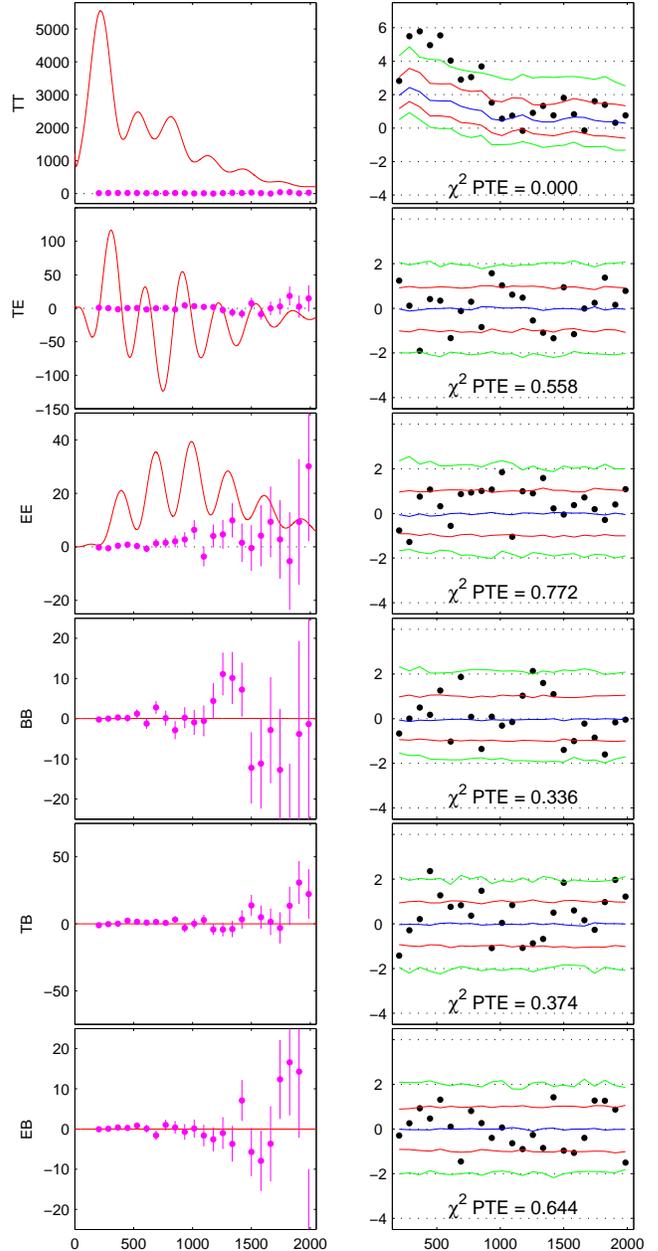}}
\caption{{\it Left:} Frequency difference spectra and {\it Right:}
the associated bandpower deviations.
$\chi^2$ is taken using the bandpower covariance matrix
and converted to the PTE value shown on each panel.
(For further explanation see the captions to the analogous
Figures~\ref{fig:spec_res} and~\ref{fig:bpdevs}.)}
\label{fig:bpdevs_fjack}
\end{figure}

We next take power spectra of the frequency difference maps and compare
them to simulation as was done for the jackknifes in Section~\ref{sec:jackknifes}.
Figure~\ref{fig:bpdevs_fjack} shows the resulting bandpower
deviations --- PTE values analogous to those in Table~\ref{tab:ptes}
are shown in the Figure.
The $TT$ spectrum shows larger deviations at low multipoles
than expected from the simulations, and the probability that
these are caused by the differing ``tiling'' effect alone
is low.
However the absolute value of these bandpowers is
$\approx 15$~$\mu$K$^2$ to be compared to the 1000's of $\mu$K$^2$
in the un-differenced map --- i.e.\ the fractional cancellation failure
is very small, and completely irrelevant compared to the sample
variance in the $TT$ spectra.
It is not clear whether the excess cancellation failure is
due to instrumental effects or real foreground signal.

\subsection{Predicted diffuse foreground levels}

Our field was chosen to partially overlap the B03 deep field to allow
absolute calibration against that map.
They in turn chose the location based in part on the position
of the Sun during their balloon flight.
Although low, the foreground emission in this region is
not the lowest available on the sky.

To obtain estimates for the expected level of foreground
dust emission we use the FDS model 8 extrapolation of IRAS
maps~\citep{finkbeiner99} --- this is shown in
Figure~\ref{fig:fieldmap}.
For synchrotron emission we use an extrapolation
of 408~MHz maps~\citep{haslam81,finkbeiner01}.
We pass these maps through the QUaD simulation pipeline,
including the field differencing and filtering operations.
For dust the maximum of the resulting $TT$ spectra is
$\sim4$~$\mu$K$^2$ (150~GHz, $\ell=200$) while the synchrotron
maximum is a negligible 0.03~$\mu$K$^2$
(100~GHz, $\ell=200$).
Although these models are possibly not the current
best available data it is clear that such extrapolations
will not give detectable levels in either temperature
or polarization, in the presence of CMB, and given the
sensitivity of QUaD.

\subsection{Point Sources}

Three discrete sources are visible in the 100 and 150~GHz $T$
maps shown in Figure~\ref{fig:sigmaps}.
As described in Section~\ref{sec:apmask} these are masked
before taking the power spectra.
Turning off the masking of these sources we find that
the 100~GHz $TT$ bandpowers increase by $\approx10$\% at
the highest multipole considered in this paper (2000)
with the increase falling off to lower multipoles.
The 150~GHz $TT$ spectrum shows an increase of $\approx3$\%
at $\ell$ of 2000 and a similar falloff.
The $EE$ and $BB$ bandpowers show changes of $\ll 1$\%.

The source flux distribution $dN/ds$ is typically a power
law with the majority of the anisotropic power being
contributed by the brightest few sources.
We therefore estimate the residual point source contribution
to be $<3$\% in the highest bin of 100~GHz $TT$, $<1$\%
in the highest bin of 150~GHz $TT$ and negligible
in all other spectra.

\subsection{Template cross correlation}

To test for the possibility of emission correlated with
thermal dust, but stronger than expected on the basis of
extrapolation, we have carried out a template cross correlation
study.
After passing the FDS dust maps through our  
pipeline the resulting maps were cross correlated with the  
corresponding QUaD CMB maps.
Though we might expect some non-zero correlation simply by a chance
alignment of large scale structure~\citep{chiang07}, we find that  
compared to simulations there is no evidence of contamination.
Indeed analyzing the cross 
power spectra at both 100 and 150~GHz between dust and CMB on a  
per bandpower basis, against simulations, reveals no problems, nor any  
suggestion that dust foregrounds are responsible for the cancellation  
failure seen in the lower bandpowers of the frequency difference $TT$  
spectrum.

\section{Combined Spectra}
\label{sec:combspec}

\begin{figure}[h]
\resizebox{\columnwidth}{!}{\includegraphics{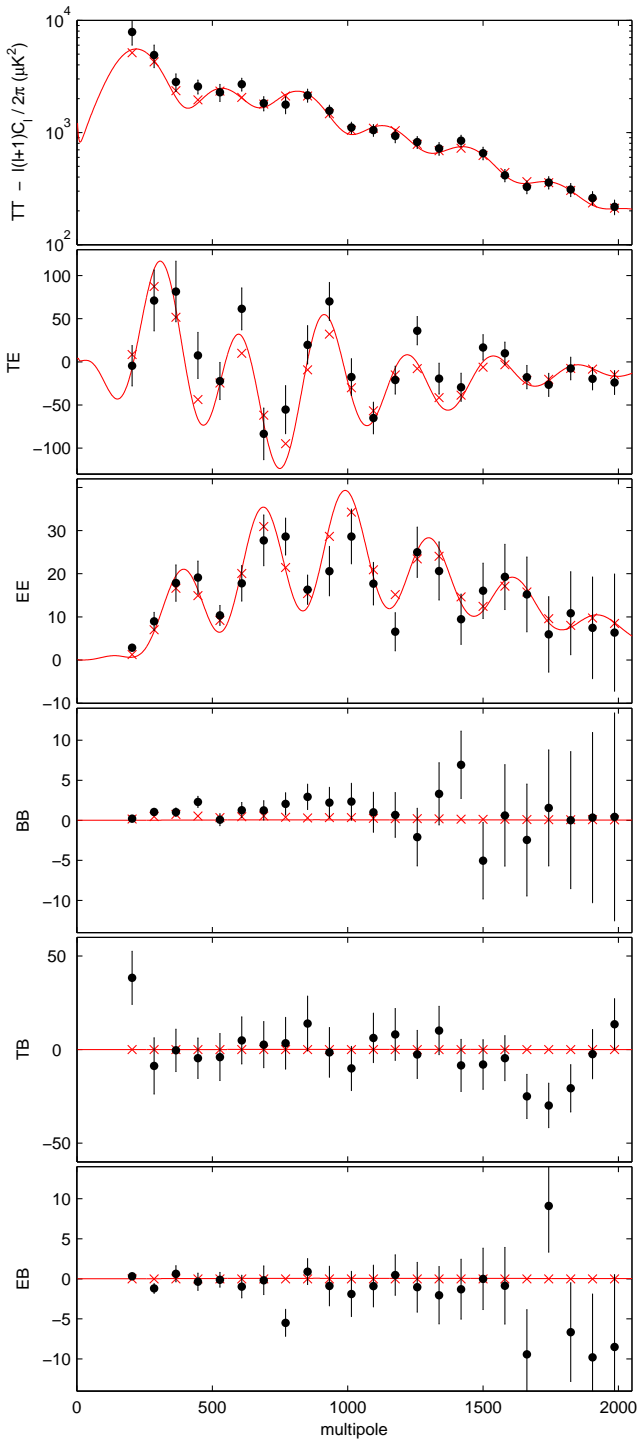}}
\caption{Combined QUaD power spectra shown as black points
with error bars.
The red crosses are the expectation values for each bandpower
given the LCDM model plotted as a red line.
Note the differing y-axis scales.}
\label{fig:spec_comb}
\end{figure}

We now wish to form a single combined set of spectra from
the 100~GHz, 150~GHz and frequency-cross results presented above.
For each bandpower of each spectrum we take the
$3 \times 3$ covariance matrix over the ensemble of
signal plus noise simulations ($4 \times 4$ for $TE$, $TB$ and $EB$).
The combination weights are the column (or row) sums of the inverse of
this matrix.
The improvement over the 150~GHz bandpower uncertainties is
between zero and 30\% depending on whether the bandpower
is signal or noise dominated.
The BPWF's are also combined.
Figure~\ref{fig:spec_comb} shows the combined spectra as
compared to their expectation values under LCDM.
The plotted bandpowers, together with their covariance matrices
and bandpower window functions, are
available in numerical form at http://quad.uchicago.edu/quad.
In contrast to the jackknife spectra, for the signal spectra
we find the simulated bandpower distributions to be Gaussian
(Cf.\ Section~\ref{sec:jackknifes}).

\subsection{Absolute Calibration and Beam Systematics}
\label{sec:syserr}

In addition to the sample and noise variance error bars
shown in Figure~\ref{fig:spec_comb} there are two
major sources of systematic uncertainty.
The first of these is the uncertainty on our absolute
calibration against the B03 maps.
As mentioned in Section~\ref{sec:abscal} we
estimate this uncertainty as 5\% in temperature units
(10\% in power).
To estimate this uncertainty one simply
multiplies the bandpower expectation values by 0.1, and takes
the outer product as an addition to the bandpower
covariance matrix:

\begin{equation}
\mathbf{C}_a = \sigma_a^2 \mathbf{e}^t \mathbf{e}
\end{equation}

\noindent where $\sigma_a=0.1$.

The second major systematic effect is uncertainty on
the width of the telescope beam.
It is highly unlikely that the widths used in the
simulations are significantly broader than the true values
--- they are very close to the results of physical
optics calculations~\citep{osullivan08}.
However, it is conceivable that we have somehow under-estimated
the pointing wander and associated effects discussed in
Section~\ref{sec:sigsims} above, causing the effective
overall beam width in the simulations to be narrower
than that in the real maps, and the suppression factor curve
plotted in Figure~\ref{fig:filtbeam} to be higher than it
should be.
This would cause the corrected bandpower values
to be biased low with increasing $\ell$.

In addition we have not yet carried out exhaustive
investigations of the beam shape and measurement uncertainties.
Therefore we very conservatively assign a beam width
uncertainty of 10\% for this analysis with the expectation
that this will be improved upon in a future analysis dedicated
to high $\ell$ $TT$.
The effective beam FWHM for the combined spectra is
$4.1'$ and the fractional shift in the bandpower values
which would result from increasing this is well approximated by

\begin{equation}
S_\ell = \frac{W_\ell}{W'_\ell} - 1 = e^{\sigma_b^2 (\delta^2+2\delta) \ell(\ell+1)} - 1
\end{equation}

\noindent where $\sigma_b = \theta_\mathrm{FWHM} / \sqrt{8 \ln 2}$, and
$\delta=0.1$ is the fractional beam error.
To estimate this uncertainty one multiplies the bandpower expectation values
by the $S_\ell$ factors calculated at the band center $\ell$ values,
and takes the outer product as an addition to the bandpower
covariance matrix.

\begin{equation}
\mathbf{C}_w = \left( \mathbf{e} S_\ell \right)^t \left( \mathbf{e} S_\ell \right)
\end{equation}

\begin{figure}[h]
\resizebox{\columnwidth}{!}{\includegraphics{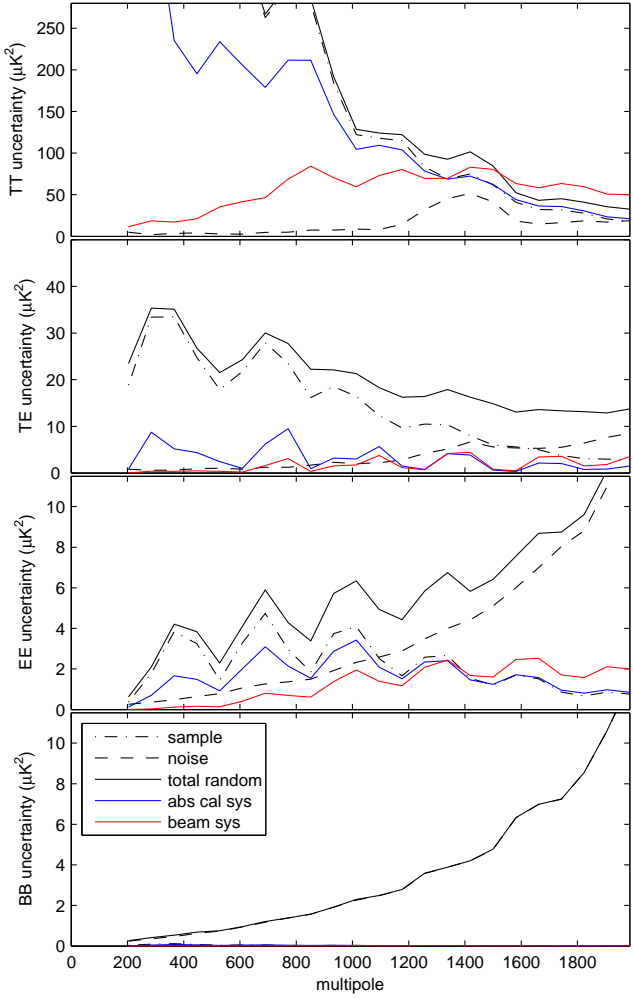}}
\caption{Contributions of the various sources of uncertainty
for the combined bandpowers.}
\label{fig:comb_uncers}
\end{figure}

Figure~\ref{fig:comb_uncers} shows the magnitude of these
uncertainties, as well as the sample and noise variance,
for each of our combined spectra.
All spectra except $BB$ are sample variance dominated at
lower $\ell$, and all but $TT$ are noise dominated at high
$\ell$.
Absolute calibration uncertainty is sub-dominant to random
uncertainty for all spectra at all $\ell$.
Beam uncertainty becomes the dominant effect for $TT$
at high $\ell$. 

\subsection{Comparison to LCDM}
\label{sec:comp2lcdm}

\begin{figure}[h]
\resizebox{\columnwidth}{!}{\includegraphics{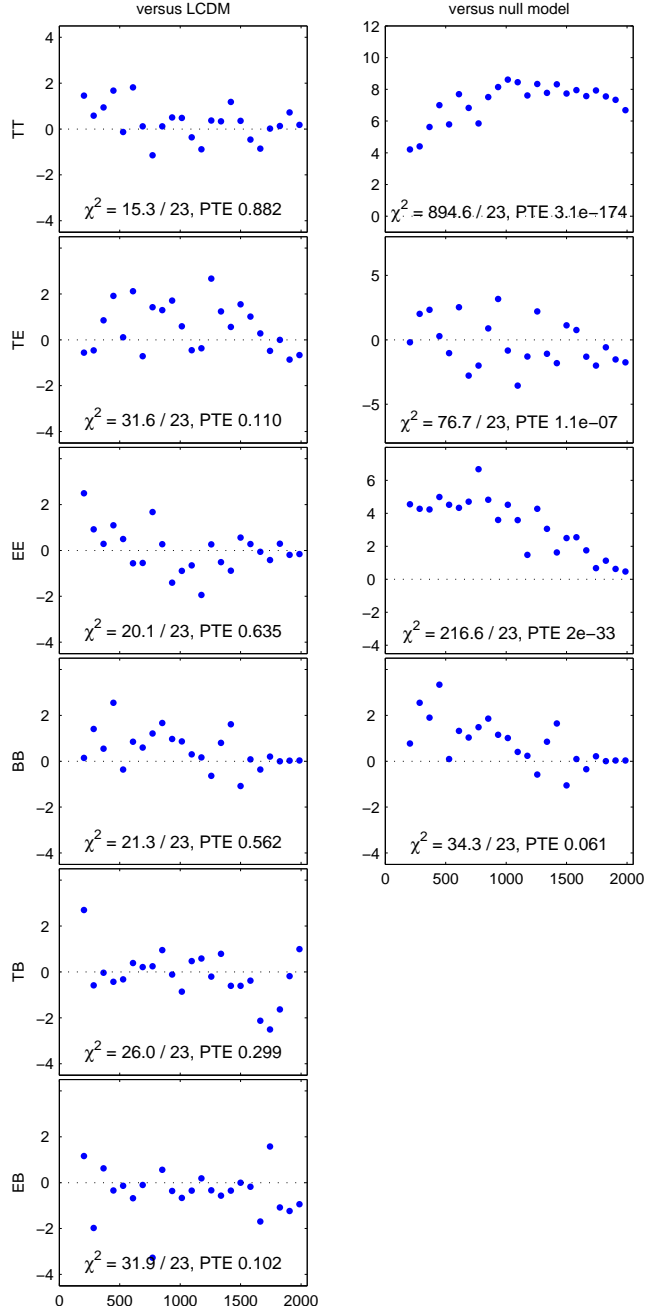}}
\caption{Bandpower deviations comparing the combined
spectra to {\it left:} LCDM, and {\it right:} the null model.
The horizontal axis is multipole number.
The respective model expectation values have been subtracted from the
spectra shown in Figure~\ref{fig:spec_comb} and the result
divided by the error bars.
Noted on each panel is the $\chi^2 / \mathrm{ndf}$ and the probability
to exceed this value by chance ($\chi^2$ being calculated using
the bandpower covariance matrix).
Note the differing $y$-axis ranges in the right column.
Note also that the $BB$ LCDM expectation values contain a significant
leakage contribution and this is why the left and right $BB$
panels differ so much --- see Section~\ref{sec:ebmix} for details.}
\label{fig:bpdevs_comb}
\end{figure}

As mentioned in Section~\ref{sec:combspec}, for the signal
spectra we find the simulated bandpower distributions to be Gaussian.
In addition we find the simulated $\chi^2$ distributions to be
consistent with the analytical expectation.
Therefore, in contrast to the jackknife spectra, for the signal
spectra we quote the analytical probability
to exceed.
Figure~\ref{fig:bpdevs_comb} shows the bandpower
deviations comparing the combined spectra to
LCDM and the null model.
Our results are clearly perfectly compatible with LCDM
and crushingly incompatible with the no polarization
hypothesis.
Note that $\sim 10$ $EE$ bandpowers have $> 4 \sigma$
significance.

\subsection{Investigation of $EE$ peaks}

\begin{figure}[h]
\begin{center}
\resizebox{0.8\columnwidth}{!}{\includegraphics{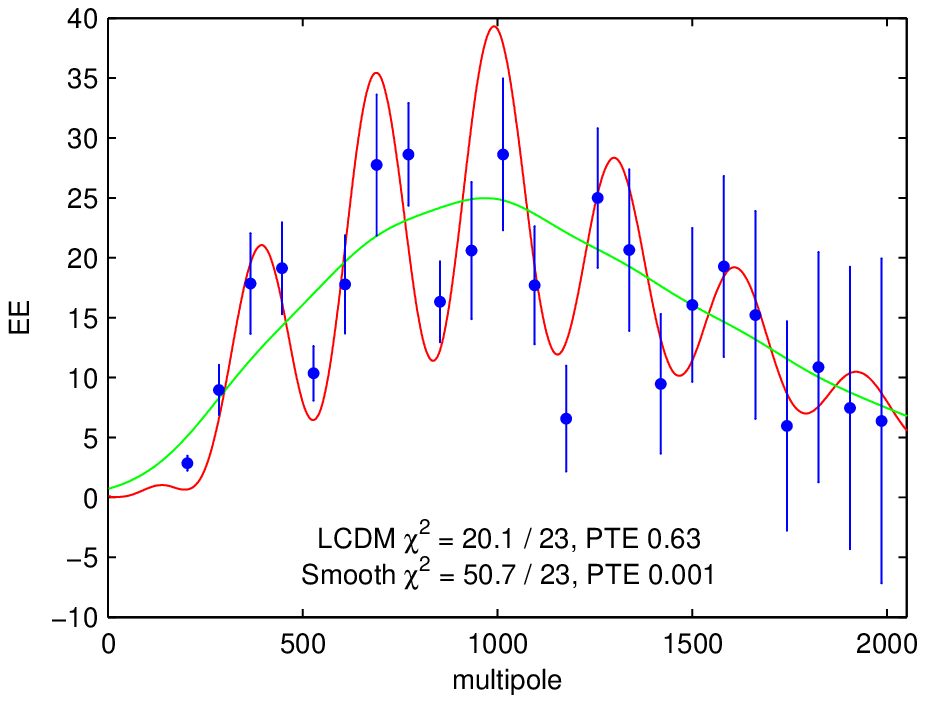}}
\end{center}
\caption{Comparing the combined $EE$ spectrum (blue points) to
LCDM (red curve) and a smooth curve without peaks (green curve).
The $\chi^2 / \mathrm{ndf}$, and the probability
to exceed this value by chance, is noted for each model
($\chi^2$ being calculated using the bandpower covariance matrix).}
\label{fig:comb_fit}
\end{figure}

Looking at Figure~\ref{fig:spec_comb} we appear to see three
or four of the expected acoustic peaks in the $EE$ spectrum.
It is interesting to ask at what significance these have been
detected.
One way to do this is to determine the $\chi^2$ of the observed
bandpowers against a version of the LCDM model which has been
smoothed sufficiently to remove the peaks.
Figure~\ref{fig:comb_fit} shows the result --- the LCDM model
has been convolved with a Gaussian with $\sigma_\ell = 150$.
We see that the probability that such a model is correct is very
low, the $\chi^2$ PTE being $0.001$.

\begin{figure}[h]
\begin{center}
\resizebox{\columnwidth}{!}{\includegraphics{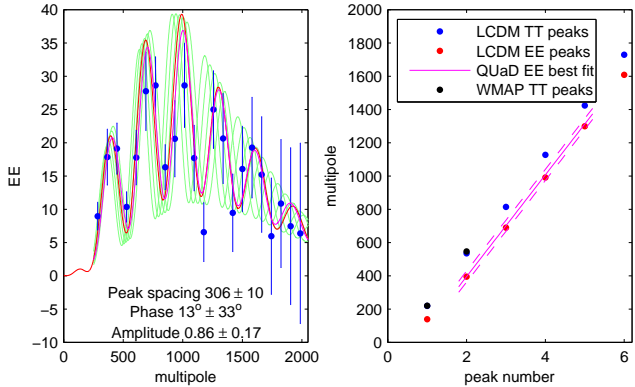}}
\end{center}
\caption{{\it Left:} Fitting a ``toy-model'' to the combined $EE$ spectrum
to determine the peak spacing, phase and amplitude, and the uncertainties
thereon.
The red line is the initial LCDM model, which is then smoothed
and sinusoidally modulated with a range of peak spacings
from 280 to 320 to generate the example family of curves shown in green.
The best fitting model curve is shown in magenta, and the
associated parameter constraints indicated.
{\it Right:} The location of the $TT$ and $EE$ peaks under
LCDM as compared to the best fit model and its uncertainty range.
The WMAP points are from~\cite{page03}.}
\label{fig:eepeakphase}
\end{figure}

In LCDM the $TT$ and $EE$ peaks are approximately half a
cycle out of phase with one another.
If $\ell_s$ is the peak spacing, and $n$ is the peak
number (starting from one), then the approximate locations of the
$TT$ peaks are $(n-1/4)\ell_s$, while the $EE$ peaks are at $(n-3/4)\ell_s$.
To investigate how well our $EE$ bandpowers constrain
the peak spacing, phase and amplitude we carry out an analysis
similar to that of~\cite{readhead04}.
Subtracting the smoothed version of the LCDM $EE$ spectrum
from the un-smoothed (i.e.\ subtracting the green from the red
curves in Figure~\ref{fig:comb_fit}) results in a series
of approximately sinusoidal modulations whose fractional
amplitude dies away close to linearly from the 2nd to the 9th
peaks.
This fact allows us to generate ``toy-model'' $EE$ spectra
which follow the envelope of LCDM using

\begin{equation}
t(\ell) = s(\ell) + a \: v(\ell) \: \sin\left(2\pi\frac{\ell}{\ell_s} + p \right)
\end{equation}

\noindent where $s(\ell)$ is the smoothed version of the LCDM spectrum,
$v(\ell)$ is the linear falloff, $a$ is a re-scaling of
the amplitude, and $p$ is the phase.
We then fit this model to the data ---
Figure~\ref{fig:eepeakphase} shows the results.
We find $\ell_s = 306 \pm 10$, $p = 13^\circ \pm 33^\circ$
and $a = 0.86 \pm 0.17$, consistent with LCDM,
as is shown in the right part of the figure.
Using WMAP $TT$ data \cite{page03} found the spacing
between the first and second peaks to be $315\pm2$.
The consistency of peak phases
and spacings between temperature data and these new QUaD
$EE$ results constitutes yet another confirmation
of the acoustic oscillation paradigm of CMB anisotropies.
\cite{readhead04} allowed only the phase and amplitude
to be free parameters finding $p=24\pm33^\circ$ ---
making this restriction we find $p=9\pm13^\circ$.

\subsection{$BB$ limits and comparison to other experiments}

\begin{figure*}[h]
\resizebox{!}{0.8\textheight}{\includegraphics{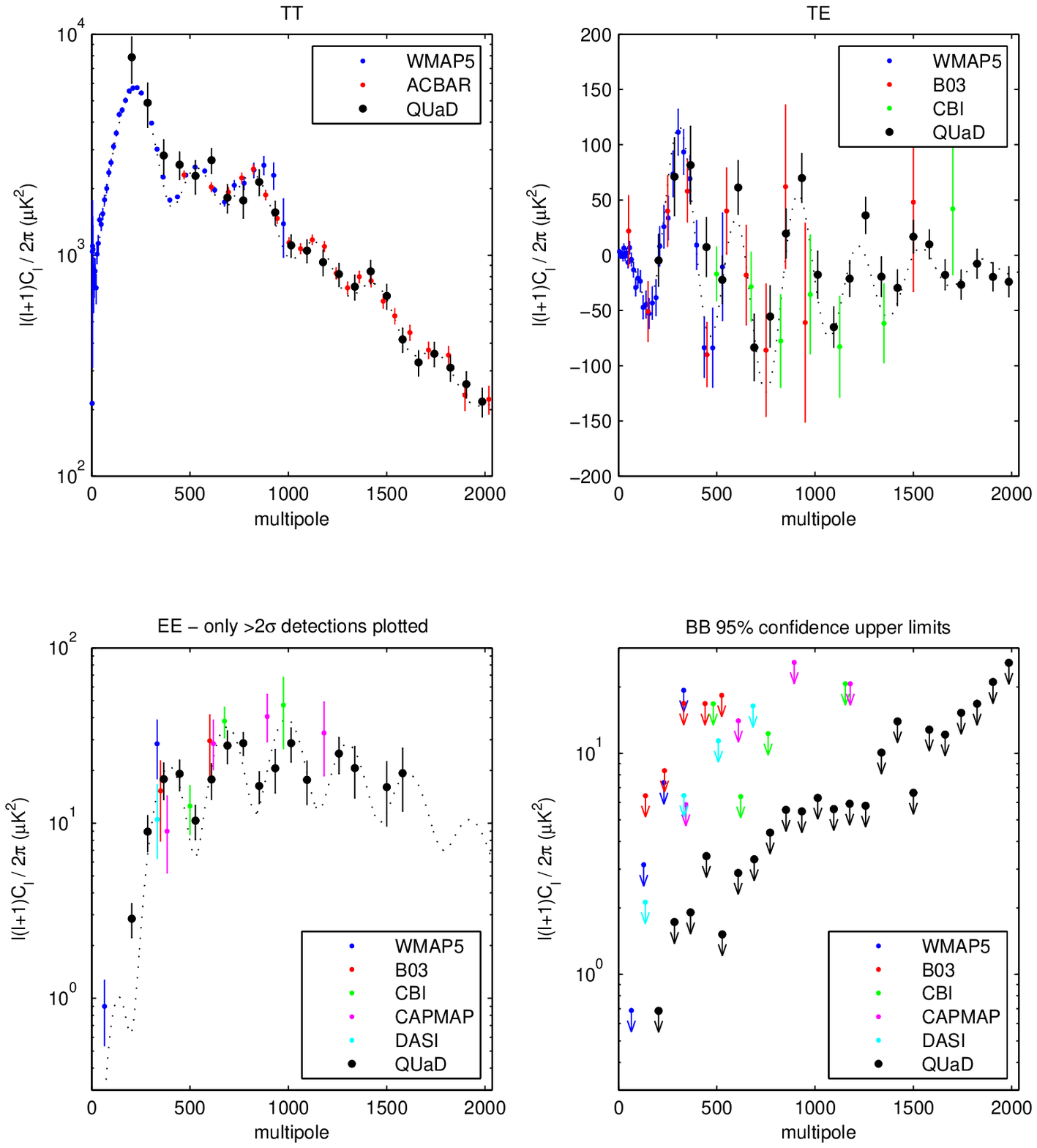}}
\caption{QUaD power spectra compared to results from
WMAP~\citep{nolta08}, ACBAR~\citep{reichardt08},
B03~\citep{piacentini06,montroy06}, CBI~\citep{sievers05},
CAPMAP~\citep{bischoff08} and DASI~\citep{leitch05}.
The $BB$ upper limits are stated values where provided,
and otherwise the 95\% point of the positive part of the
bandpower pdf.}
\label{fig:spec_comp}
\end{figure*}

The $BB$ results shown in Figure~\ref{fig:spec_comb}
are consistent with zero sky power, and 
we therefore interpret these results as upper limits.
To convert the observed values into confidence
limits we find the 95\% integral point of the positive
part of the bandpower probability density function (which is assumed to be
Gaussian with mean and spread as in Figure~\ref{fig:spec_comb}).
Figure~\ref{fig:spec_comp} shows a comparison of these
limits, and our other spectra, to published results from other experiments.
In this figure $EE$ is shown on a log-scale, and hence to avoid
clutter only bandpowers whose center value is more than twice
the distance between the center value and the lower end of
the 68\% confidence limit are shown.
For $TE$, $EE$ and $BB$ QUaD breaks new ground --- for $TT$
the high $\ell$ precision is comparable to ACBAR,
although the beam uncertainty is larger in the present analysis.

\subsection{$E$ and $B$ maps}

\begin{figure}[h]
\begin{center}
\resizebox{\columnwidth}{!}{\includegraphics{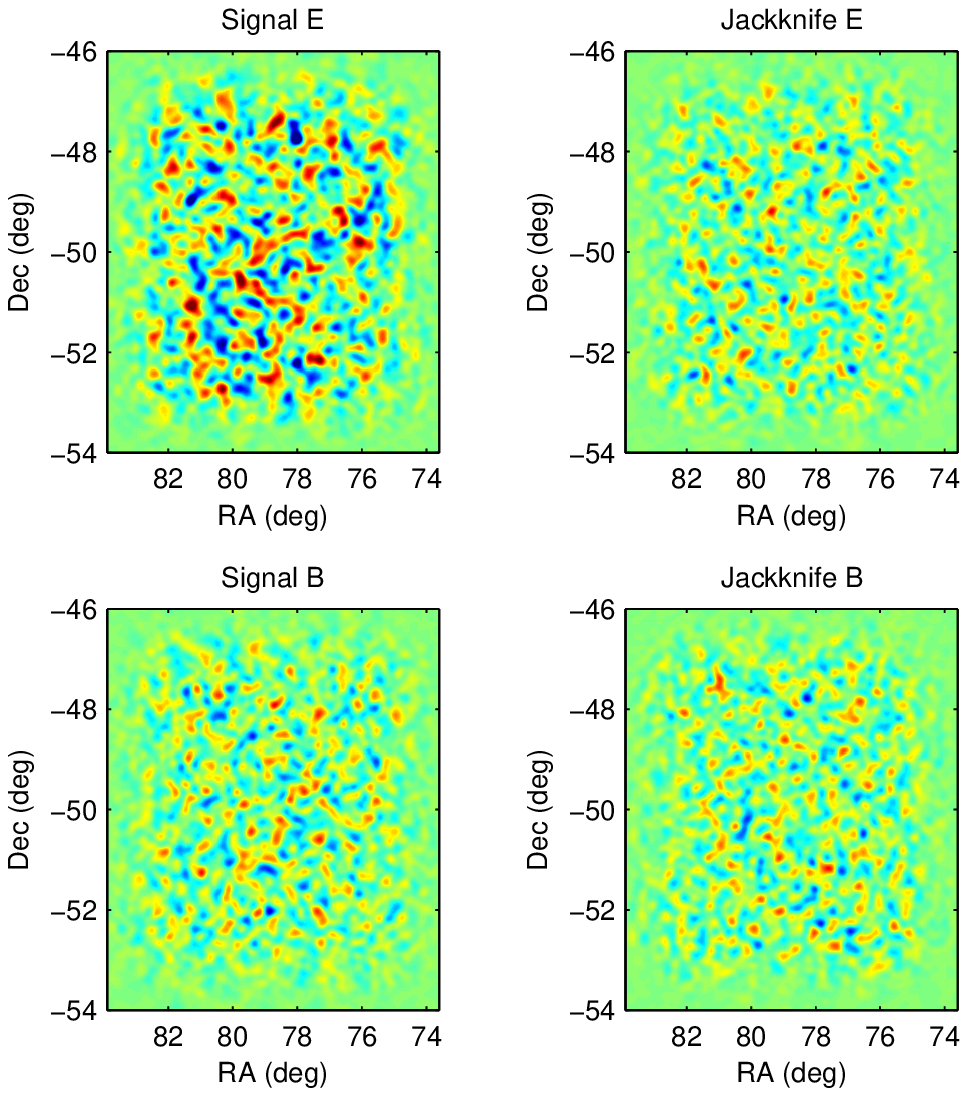}}
\end{center}
\caption{$E$ and $B$ signal and (deck) jackknife maps.
The color scale is $\pm20$~$\mu$K.
These maps have been apodized and filtered to enhance signal-to-noise
--- see text for details.}
\label{fig:ebmap}
\end{figure}

In Figure~\ref{fig:spec_comb} it is clear that we detect
dramatically more $E$-mode power on the sky than $B$-mode.
Another way to visualize this is in the image plane.
Having converted the apodized $Q$ and $U$ maps to $E$ and $B$ Fourier modes
as described in Section~\ref{sec:maps2spec}
we can take the inverse Fourier transform to produce $E$ and $B$ maps.
To enhance the signal to noise we apply a Fourier space filter\footnote{
Here we use a Weiner filter assuming the LCDM $EE$ spectrum,
but a band pass filter $200<\ell<2000$ produces
a qualitatively similar result.}.
Figure~\ref{fig:ebmap} shows the result --- the $E$ map
contains far more structure than the $B$ map.
To confirm that the residual $B$ signal is consistent
with noise the figure also shows equivalent deck
jackknife maps.

\section{Systematic Issues}
\label{sec:systematics}

There are a wide range of systematic effects which potentially
mix $T$ into pol.\ and/or $EE$ into $BB$.
In practice, as seen in Section~\ref{sec:comp2lcdm}, our
results are consistent with the LCDM prediction
of lensing $BB$, which is an order of magnitude
smaller than our noise induced bandpower uncertainty,
and hence effectively zero for the purposes
of our experiment.

While there are many ways to produce false $BB$ signal,
it is virtually impossible that true $BB$ power could
somehow be canceled out through systematic effects.
The fact that our $BB$ results are consistent with
zero is therefore powerful evidence that systematic mixing effects have
been controlled to the required level of precision.

Note that from an instrumental point of view this
is a fortunate accident of cosmology --- the
theoretical prediction is that the
CMB sky presents us with a high purity
``test pattern'' against which we can validate our experiment.
If we did see $BB$ power at a level greater than the LCDM prediction
extensive investigation of possible systematics would
be required to attempt to verify that it was real.
But since we do not see any $BB$ further investigation
of mixing systematics is arguably unnecessary.
However, for completeness, we present some additional discussion
below.

\subsection{$EE$ to $BB$ mixing due to the sky cut}
\label{sec:ebmix}

Since the transform from $Q,U$ to $E,B$ is non-local,
for less than full sky coverage leakage occurs
from the $EE$ spectrum to $BB$ (and vice versa, although
this is irrelevant under LCDM).
In this analysis we deal with this effect through the
cross-spectra BPWF's discussed in Section~\ref{sec:bpwf}.
For LCDM our $BB$ expectation values peak at 0.7~$\mu$K$^2$
for $\ell=360$ and fall rapidly to higher multipoles.
At the current level of sensitivity including
the cross spectral BPWF's is just starting to become necessary ---
as we see from the increased $\chi^2$ value
going from the LCDM to null model $BB$ panels in
Figure~\ref{fig:bpdevs_comb}.

Our signal only simulations include several other
effects which potentially produce false $BB$ signal
(including beam centroid mismatch and grid angle uncertainty --- see
Section~\ref{sec:sigsims}).
However for the standard simulation parameters we
find that the sky cut effect is dominant since
the mean $BB$ simulated spectrum follows the expectation
values calculated using the cross-spectral BPWF's.

The original MASTER approach~\citep{hivon02} includes a ``de-mixing''
correction, and this was subsequently extended to
polarization (e.g.~\cite{brown05}).
Such techniques reduce spectral mixing in the mean,
and hence also reduce the cross-spectra BPWF's.
More recently a technique was proposed by~\cite{smith07} 
which results in much lower mixing from $E$ to $B$ within
any given realization --- 
we have implemented this in the flat sky case, and confirmed
with simulations that it works.
However since this complication is unnecessary we
do not include it in this paper.

\subsection{Curved sky versus flat sky}

The analysis presented here uses flat sky 
power spectrum estimation while
most recent CMB analyses have been conducted in the 
spherical harmonic basis.
By using full curved sky maps as the input to our 
signal timestream simulation,
our power estimation is normalized to recover 
the curved-sky input power.
In addition these simulations empirically test
for problems associated with 
these flat-sky estimators, which are found to be
negligible for our small ($\sim 6 \times 6^\circ$)
patch of sky.
As mentioned above $EE$ to $BB$ leakage is dominated
by the cut sky effect and is well reproduced in the simulations
by the cross-spectral BPWF's
--- for QUAD the use of flat-sky power estimators does not contribute 
significantly to this leakage.

\subsection{Relative Gain Calibration}
\label{sec:ditherrelgains}

In Section~\ref{sec:relgains} the elevation nod based method we
use to equalize the detector gains was described.
This method appears to be very accurate, and we do not include
any scatter in the standard simulations.
However, since error in the pair relative gains
leads to $T$ to pol.\ leakage
there are a couple of issues which one might worry about.

The elevation nods integrate over the full
beam including any far sidelobes, whereas when mapping the
CMB, it is the ratio of the main lobe gains which we wish
to know.
Sky dips extending over a much larger zenith
angle range (5 to 45$^\circ$) follow the expected
$\sec(\theta)$ dependence very closely ($<1$\% rms residual), indicating that
the elevation nods are not distorted by sidelobe response.

The two detectors of each pair share a common feed horn
and filters, but one might worry that the bandpasses of
the fore and aft detectors might still differ.
Since the atmospheric emission has a different frequency
spectrum from the CMB this might lead to a systematic error
in the relative gain within each pair.

To test how such errors would play out in practice
we ran some special simulations where the input
sky maps contain $T$ only, and where the detector
pair gains are deliberately systematically distorted such that
$g_\mathrm{fore}/g_\mathrm{aft}=1.03$ --- far worse than the
$<1$\% constraint on possible mismatch which we derive
from RCW38 observations.
We find that the resulting $EE$ and $BB$ spectra
follow the same form as the input $TT$ spectrum
with a peak of 0.4~$\mu$K$^2$ at $\ell=200$.
It is important to note that even for a systematic
error like this there is still considerable averaging
down as pairs of different angles, at different
telescope orientations, leak $T$ into, for
instance, both $+Q$ and $-Q$ within a given map pixel.
Making the gain ratio errors a random 3\% across the focal
plane (but fixed over time) the averaging down
is much more effective and the peak leakage
becomes 0.1~$\mu$K$^2$.

\subsection{Pair beam mismatch}
\label{sec:beammismatch}

As mentioned in Section~\ref{sec:sigsims} we observe
repeatable beam centroid offsets between the two
halves of each detector pair, fixed in the instrument frame,
with rms magnitude of $\sim 0.1'$.
$T$ to pol.\ leakage introduced by this effect is included
in the standard simulations but makes a negligible
contribution.
However for interest we run some special simulations with only
$T$ input introducing random pair centroid offsets with rms magnitude of
$2'$ --- twenty times the observed value.
We find that the resulting leakage has a broad peak at
$\sim2$~$\mu$K$^2$ around $\ell\sim1000$ for 100~GHz, and
$\sim1$~$\mu$K$^2$ for 150~GHz.
In addition we run simulations under the totally unrealistic
scenario that all pairs are systematically offset in the
same direction by $1'$ and find leakage
of 6~$\mu$K$^2$ at $\ell\sim1000$.

For the standard simulations we use the individual
channel major/minor fit widths and orientation
angles as measured using PKS0537-441.
Some of the apparent variation between channels in
these observations is measurement noise, but there is
some real variation, and $\leq 10$\% elongation.
To investigate the impact of differential
beam size we run special $T$ only simulations with
both the major and minor axis FWHM for one half of each
pair systematically inflated by $1.4'$ ---
the resulting leakage peaks at $\sim1.5$~$\mu$K$^2$
above $\ell\sim1500$.
For differential elongation, where we inflate only
one axis of one half of each pair
(with a common orientation angle for all pairs), we find
$\sim1$~$\mu$K$^2$ above $\ell\sim1500$.

\subsection{Polarization angle}
\label{sec:polang}

As seen in Section~\ref{sec:comp2lcdm} our results
are consistent with LCDM ---
we detect considerable $EE$ power and no $BB$.
This is in itself confirmation that the assumed
polarization angles of the detectors are known to sufficient
accuracy.
If we did see significant $BB$ we might suspect
that it was false signal due to incorrect angles.
But it would be nearly impossible for
a sky which truly had $B$-mode power to appear
not to due to incorrect detector angles.

To confirm this we re-generated the real
maps using detector angles systematically biased
from the best estimate values
by far more than the 1$^\circ$ estimated uncertainty
(see Section~\ref{sec:polpar} and the Instrument Paper).
For a 5$^\circ$ bias there is almost no effect --- the $\chi^2$
PTE versus LCDM for the $BB$ spectrum (as shown in
Figure~\ref{fig:bpdevs_comb}) falls from 0.56 to 0.21.
Only at 10$^\circ$ do we start to see significant extra
signal in the $BB$ bandpowers, and $\chi^2$ failure
versus LCDM for $BB$ (and $EB$).

\subsection{Polarization efficiency}
\label{sec:poleff}

As mentioned in Section~\ref{sec:polpar}, and
our Instrument Paper, we
measure the polarization efficiency of our detectors
to be $\epsilon=0.08$ with rms scatter of 0.015.
An additional calibration factor $\gamma=(1-\epsilon)/(1+\epsilon)$
is then applied to the pair difference data.
Random errors in $\epsilon$ will average down, while a
systematic error will translate into a shift in the
absolute calibration of the polarization power spectra
$\sim 4$ times as large (including the additional doubling
when going from units of temperature to power).
Since we estimate the uncertainty on $\bar{\epsilon}$
to be~$<0.02$ the implied uncertainty
on the polarization spectra is sub-dominant
to the overall absolute calibration uncertainty.

\subsection{Moon pickup}
\label{sec:mooncontam}

QUaD has a variety of far sidelobes as described in the
accompanying Instrument Paper.
It is not clear which of these produces the bulk of
the highly polarized ground pickup mentioned in
Section~\ref{sec:fielddiff} --- however since the field
differencing is so effective at removing this contamination
this is probably of academic interest only.
However, any source of contamination which moves with
respect to the ground
will not be removed by field differencing.

Part of our basic low level data reduction infrastructure
involves making single pair sum and difference maps
for each 8 hour block of observations.
On certain days when the Moon is high above the horizon,
but at a very large angle from the telescope pointing direction,
we see obvious stripes in these maps.
Even after cutting these visibly contaminated days from the 
analysis, we
saw a strong peak towards zero in the PTE distribution
as show in Figure~\ref{fig:pte_dist}, and some
$\chi^2$ values were much too large.

Extensive effort has been required to elucidate the coupling
mechanism by which the Moon enters the CMB field data.
We have determined that radiation reflects off the
inside surface of the foam cone creating a polarized ring
sidelobe at $\approx100^\circ$ from the main beam.
The shape of the pickup across any given scan
has been successfully modeled using the position of the Moon
relative to the telescope pointing direction, the feed offset
angle, and the detector polarization angle (see the
Instrument Paper for details).

Using this model we cut data periods 
which are potentially contaminated by the Moon.
For this analysis the cut is simple and quite aggressive
--- for any day where the Moon passes within a generous band
about the ring sidelobe we simply discard the entire day.
A future analysis could retain slightly more data
by selectively cutting channels, and using a time granularity
shorter than a whole day.
Performing this cut we reject an additional 59 days of observation
but as seen in Section~\ref{sec:jackknifes} the jackknife tests
pass, meaning that we can
be confident of the final power spectrum results.

\section{Conclusions}
\label{sec:conclusions}

We have described the observations, data reduction, simulation
and power spectrum analysis of the QUaD experiment.
The results reported here are from 143 days of data taken in the
second and third (final) seasons of observation, employing
a conservative Moon cut, and simple lead-trail differencing.
A future analysis may be able to include more data, and/or
reduce the information lost in the ground removal.

The three sets of power spectra, 100~GHz, 150~GHz and
frequency-cross, have been subjected to an extensive set of
jackknife tests and residual systematic contamination has been
shown to be undetectable above the instrumental noise.

The combined spectra improve very considerably
in sensitivity over previous results, and are consistent with LCDM ---
the standard cosmological model has passed yet another
stringent test.
Furthermore we find that a smooth curve is a very poor
fit to the observed $EE$ spectrum --- acoustic peaks
in the $EE$ spectrum have been detected with high significance
for the first time.
The impact of possible instrumental systematics has been considered in
detail, but in fact the tight upper limits on $BB$ power
obtained are in themselves extremely powerful evidence
that such effects are adequately controlled.

\acknowledgements

QUaD is funded by the National Science Foundation in the USA, through
grants AST-0096778, ANT-0338138, ANT-0338335 \& ANT-0338238, by the
Particle Physics and Astronomy Research Council in the UK, and by the
Science Foundation Ireland. We would like to thank the staff of the
Amundsen-Scott South Pole Station and all involved in the United
States Antarctic Program for the superb support operation which makes
the science presented here possible. Special thanks go to our intrepid
winter over scientist Robert Schwarz who spent three consecutive
winter seasons tending the QUaD experiment. The BOOMERanG
collaboration kindly allowed the use of their CMB maps for our
calibration purposes.  MLB acknowledges the award of a PPARC
fellowship. SEC acknowledges support from a Stanford Terman
Fellowship. JRH acknowledges the support of an NSF Graduate Research
Fellowship and a Stanford Graduate Fellowship. CP and JEC acknowledge
partial support from the Kavli Institute for Cosmological Physics
through the grant NSF PHY-0114422.  EYW acknowledges receipt of an
NDSEG fellowship.
JMK acknowledges support from a John B.\ and Nelly L.\ Kilroy Foundation 
Fellowship.

\bibliographystyle{apj}
\bibliography{ms}

\end{document}

%% file: ptetab.tex
Deck angle\\
TT  & 0.102  & 0.072  & 0.030 \\
TE  & 0.066  & 0.040  & 0.174  & 0.586 \\
EE  & 0.874  & 0.624  & 0.210 \\
BB  & 0.990  & 0.316  & 0.946 \\
TB  & 0.506  & 0.410  & 0.520  & 0.632 \\
EB  & 0.736  & 0.210  & 0.952  & 0.180 \\
\\
Scan direction\\
TT  & 0.650  & 0.238  & 0.982 \\
TE  & 0.316  & 0.444  & 0.870  & 0.718 \\
EE  & 0.656  & 0.490  & 0.372 \\
BB  & 0.882  & 0.156  & 0.982 \\
TB  & 0.274  & 0.430  & 0.154  & 0.254 \\
EB  & 0.540  & 0.332  & 0.826  & 0.418 \\
\\
Split season
\\
TT  & 0.202  & 0.020  & 0.408 \\
TE  & 0.962  & 0.296  & 0.300  & 0.472 \\
EE  & 0.116  & 0.038  & 0.066 \\
BB  & 0.896  & 0.036  & 0.100 \\
TB  & 0.946  & 0.494  & 0.896  & 0.166 \\
EB  & 0.860  & 0.612  & 0.972  & 0.890 \\
\\
Focal plane\\
TT  & 0.352  & 0.472  & 0.232 \\
TE  & 0.858  & 0.872  & 0.448  & 0.936 \\
EE  & 0.654  & 0.174  & 0.944 \\
BB  & 0.696  & 0.642  & 0.708 \\
TB  & 0.888  & 0.422  & 0.008  & 0.456 \\
EB  & 0.932  & 0.640  & 0.354  & 0.344 \\